# Energy, Momentum and Angular Momentum Transfer Between Electrons and Nuclei

Chen Li,[1,2,3,*] Ryan Requist,[1,2] and E. K. U. Gross[1,2]

[1] *Max Planck Institute of Microstructure Physics, Weinberg 2, 06120, Halle, Germany*
[2] *Fritz Haber Center for Molecular Dynamics, Institute of Chemistry, The Hebrew University of Jerusalem, Jerusalem 91904 Israel*
[3] *Beijing National Laboratory for Molecular Sciences, College of Chemistry and Molecular Engineering, Peking University, Beijing 100871, China*

The recently developed exact factorization approach condenses all electronic effects on the nuclear subsystem into scalar and vector potentials that appear in an effective time dependent Schrödinger equation. Starting from this equation, we derive subsystem Ehrenfest identities characterizing the energy, momentum and angular momentum transfer between electrons and nuclei. An effective electromagnetic force operator induced by the electromagnetic field corresponding to the effective scalar and vector potentials appears in all three identities. The effective magnetic field has two components that can be identified with the Berry curvature calculated with (a) different cartesian coordinates of the same nucleus and (b) arbitrary cartesian coordinates of two different nuclei. (a) has a classical interpretation as the induced magnetic field felt by the nucleus, while (b) has no classical analog. Subsystem Ehrenfest identities are ideally suited for quantifying energy transfer in electron-phonon systems. With two explicit examples we demonstrate the usefulness of the new identities.

The immensity of information in the quantum mechanical wave function is an obstacle to finding a clear physical picture of microscale dynamical processes. It is thus crucial to single out a few variables that condense the most relevant information, and experience shows this is particularly successful when these variables have classical analogs. This line of thinking dates back to Ehrenfest. For a single particle described by a time dependent Schrödinger equation (TDSE), the Ehrenfest theorem bridges the quantum and classical pictures by providing equations of motion for the expectation values of position and momentum that have a strong resemblance to Newton's equations. [1]

Yet, real world systems are made up of multiple particle species. In this respect, the Ehrenfest theorem and its generalizations [2] are limited because they do not probe the multicomponent nature of the system. It would therefore be desirable to go beyond the Ehrenfest theorem in the following two ways: (i) identifying useful variables that are specific to a subsystem, and (ii) deriving their equations of motion in a form which brings to light the classical analogs they contain.

For molecules and solids, three obvious candidates for (i) are the kinetic energy, momentum and angular momentum of the nuclei. These variables are helpful in gaining insight into dynamical phenomena where energy and momentum are transferred between electrons and nuclei. For example, energy transfer is crucial for understanding the fast internal conversion of DNA and RNA [3, 4] and electronic friction-induced relaxation of molecular vibrations [5–7]. By tuning the energy transfer rate, one can control current-induced forces [8–10] in nanosystems and minimize Joule heating [11, 12]. By understanding angular momentum transfer on the microscale, one may find inspiration in designing molecular motors and refrigerators [13–16] and in studies of quantum thermodynamics [17–19].

Energy transfer in electron-phonon systems is the subject of intense and sustained research. [20] Time-resolved pump-probe spectroscopy is capable of tracking the nonequilibrium dynamics of electrons after excitation by a laser pulse. Predicting the subsequent phonon-induced electron relaxation is a challenge for existing theoretical approaches, and most work uses phenomenological models.

In this Letter, we derive Ehrenfest identities that can rigorously quantify the energy transfer between electrons and phonons. Our results in fact apply to any two-component system, though for concreteness we consider here a system of electrons and nuclei. Unlike conventional methods where quantum subsystems are described by a reduced density matrix [21, 22] or nonequilibrium Green's function [23], the essence of our approach lies in the realization that a subsystem can be treated as a pure state whose dynamics are described by an effective Schrödinger equation.

The exact factorization (EF) method [24–26] sets down the rigorous definition of a nuclear wave function that yields the exact nuclear probability density and current density. [27] We will show in this Letter that it also yields the exact nuclear angular momentum. The fact that the nuclear wave function obeys a TDSE in which all electronic effects have been condensed into scalar and vector potentials [25, 26] is the key to point (ii) above, for it is precisely this structure that allows us to identify quantities with classical analogs in the subsystem Ehrenfest identities (SEI) for the kinetic energy, momentum and angular momentum.

Let us start with the full TDSE for electrons and nuclei,

$$i\partial_t \Psi(\underline{\underline{\mathbf{r}}}, \underline{\underline{\mathbf{R}}}, t) = \hat{H}\Psi(\underline{\underline{\mathbf{r}}}, \underline{\underline{\mathbf{R}}}, t), \tag{1}$$

where $\underline{\underline{\mathbf{r}}} = (\mathbf{r}_1, \mathbf{r}_2, \cdots, \mathbf{r}_{N_e})$ and $\underline{\underline{\mathbf{R}}} = (\mathbf{R}_1, \mathbf{R}_2, \cdots, \mathbf{R}_{N_n})$ denote the electronic and nuclear coordinates, respectively. $\hat{H}$ is the full Hamiltonian which in the absence of external potentials comprises the nuclear kinetic energy $\hat{T}_n$, the electronic kinetic energy $\hat{T}_e$, electron-electron interaction $\hat{V}_{ee}$, electron-nucleus interaction $\hat{V}_{en}$, and nucleus-nucleus interaction $\hat{V}_{nn}$. The nuclear kinetic energy $T_n$, momentum $\boldsymbol{P}_n$ and angular momentum $\boldsymbol{L}_n$ are defined as the expectation values of the corresponding operators,

$$T_n = \langle\Psi| \sum_{\mu=1}^{N_n} -\frac{1}{2M_\mu}\nabla^2_{\mathbf{R}_\mu}|\Psi\rangle_{\underline{\underline{\mathbf{r}\mathbf{R}}}}, \tag{2}$$

$$\boldsymbol{P}_n = \langle\Psi| \sum_{\mu=1}^{N_n} -i\nabla_{\mathbf{R}_\mu}|\Psi\rangle_{\underline{\underline{\mathbf{r}\mathbf{R}}}}, \tag{3}$$

$$\boldsymbol{L}_n = \langle\Psi| \sum_{\mu=1}^{N_n} \mathbf{R}_\mu \times (-i\nabla_{\mathbf{R}_\mu})|\Psi\rangle_{\underline{\underline{\mathbf{r}\mathbf{R}}}}. \tag{4}$$

Here $\mu$ indexes the nuclei, $M_\mu$ are the nuclear masses, and the subscripts of the bra-kets indicate which variables are integrated over in the inner product. As a nonstationary $\Psi$ evolves, these expectation values change in time due to the coupling to the electronic subsystem.

It has been shown that $\Psi(\underline{\underline{\mathbf{r}}}, \underline{\underline{\mathbf{R}}}, t)$ can be factorized into a marginal nuclear wave function $\chi(\underline{\underline{\mathbf{R}}}, t)$ and a conditional electronic wave function $\Phi_{\underline{\underline{\mathbf{R}}}}(\underline{\underline{\mathbf{r}}}, t)$. [24–27] Furthermore, $\Phi_{\underline{\underline{\mathbf{R}}}}$ satisfies a complicated electronic equation while $\chi$ satisfies a simple nuclear TDSE, [26, 27] $i\partial_t\chi = \hat{H}_n\chi$, where $\hat{H}_n = \sum_{\mu=1}^{N_n}\frac{1}{2M_\mu}(-i\nabla_{\mathbf{R}_\mu} + \boldsymbol{A}_\mu)^2 + \epsilon$. Here $\epsilon(\underline{\underline{\mathbf{R}}}, t)$ is the scalar potential originating from the electronic equation, and $\boldsymbol{A}_\mu(\underline{\underline{\mathbf{R}}}, t) = \langle\Phi_{\underline{\underline{\mathbf{R}}}}| - i\nabla_{\mathbf{R}_\mu}|\Phi_{\underline{\underline{\mathbf{R}}}}\rangle_{\underline{\underline{\mathbf{r}}}}$ are nucleus-dependent vector potentials. By virtue of the fact that $\chi$ obeys a TDSE and can be viewed as the wave function of a closed system acted on by $\epsilon$ and $\boldsymbol{A}_\mu$, we can evaluate $T_n$, $\boldsymbol{P}_n$ and $\boldsymbol{L}_n$ equally well as

$$T_n = \langle\chi| \sum_{\mu=1}^{N_n} \frac{1}{2M_\mu}(-i\nabla_{\mathbf{R}_\mu} + \boldsymbol{A}_\mu)^2|\chi\rangle_{\underline{\underline{\mathbf{R}}}} + T_{n,geo}, \tag{5}$$

$$\boldsymbol{P}_n = \langle\chi| \sum_{\mu=1}^{N_n} (-i\nabla_{\mathbf{R}_\mu} + \boldsymbol{A}_\mu)|\chi\rangle_{\underline{\underline{\mathbf{R}}}}, \tag{6}$$

$$\boldsymbol{L}_n = \langle\chi| \sum_{\mu=1}^{N_n} \mathbf{R}_\mu \times (-i\nabla_{\mathbf{R}_\mu} + \boldsymbol{A}_\mu)|\chi\rangle_{\underline{\underline{\mathbf{R}}}}. \tag{7}$$

Here

$$T_{n,geo} = \langle\chi| \sum_{\mu=1}^{N_n} \frac{1}{2M_\mu}\Big(\langle\nabla_{\mathbf{R}_\mu}\Phi_{\underline{\underline{\mathbf{R}}}}|\nabla_{\mathbf{R}_\mu}\Phi_{\underline{\underline{\mathbf{R}}}}\rangle_{\underline{\underline{\mathbf{r}}}} - \boldsymbol{A}_\mu^2\Big)|\chi\rangle_{\underline{\underline{\mathbf{R}}}} \tag{8}$$

is an additional *geometric* term, which can be written as the tensor contraction of an inverse inertia tensor and a Riemannian metric measuring distance in the space of quantum states (see Ref. [28] and references therein.) Comparing Eqs. (5)–(7) with Eqs. (2)–(4), one can easily recognize their formal resemblance. The equivalence of Eq. (6) and Eq. (3) implies that the Ehrenfest equation for the momentum of the nuclei can be evaluated by considering either the full system or the nuclear subsystem alone, as shown in Ref. [29]. In replacing the full wave function $\Psi$ by the marginal subsystem wave function $\chi$ and the corresponding integration domain, we obtain additional terms with vector potentials $\boldsymbol{A}_\mu$ arising in conjunction with the canonical momentum operators. A similar argument applies to the nuclear angular momentum. In contrast, Eq. (5) and Eq. (2) imply that the true nuclear kinetic energy $T_n$, differs from the kinetic energy of the effective closed system described by the nuclear subsystem (marginal) TDSE, denoted as $T_{n,marg}$, by the quantity in Eq. (8) as shown in Ref. [30]. In the following, we derive the equations of motion for $T_{n,marg}$, $\boldsymbol{P}_n$ and $\boldsymbol{L}_n$, and will show that they all satisfy classical-like equations, governed by a unified force operator.

We start by applying the Heisenberg equation of motion for $T_{n,marg} = \langle\chi|\hat{t}_n|\chi\rangle$ with $\hat{t}_n = \sum_{\mu=1}^{N_n}\frac{1}{2M_\mu}(-i\nabla_{\mathbf{R}_\mu} + \boldsymbol{A}_\mu)^2$ to the nuclear TDSE, which leads to

$$\begin{aligned}\frac{dT_{n,marg}}{dt} &= i\langle\chi|[\hat{H}_n, \hat{t}_n]|\chi\rangle_{\underline{\underline{\mathbf{R}}}} + \langle\chi|\frac{\partial\hat{t}_n}{\partial t}|\chi\rangle_{\underline{\underline{\mathbf{R}}}} \\ &= \sum_{\mu=1}^{N_n}\frac{1}{M_\mu}\Big\{\langle\chi|(\partial_t\boldsymbol{A}_\mu - \nabla_{\mathbf{R}_\mu}\epsilon)\cdot(-i\nabla_{\mathbf{R}_\mu} + \boldsymbol{A}_\mu)|\chi\rangle_{\underline{\underline{\mathbf{R}}}} \\ &\quad - i\frac{1}{2}\langle\chi|\Big(\nabla_{\mathbf{R}_\mu}\cdot(\partial_t\boldsymbol{A}_\mu - \nabla_{\mathbf{R}_\mu}\epsilon)\Big)|\chi\rangle_{\underline{\underline{\mathbf{R}}}}\Big\}.\end{aligned} \tag{9}$$

Here the left hand side (LHS) of Eq. (9) is real. When we take the real part of Eq. (9), the LHS stays the same while the second term in the braces of the RHS vanishes since it is purely imaginary. Then, by introducing a velocity operator $\hat{\boldsymbol{v}}_\mu \equiv \frac{1}{M_\mu}(-i\nabla_{\mathbf{R}_\mu} + \boldsymbol{A}_\mu)$ for each nucleus and defining the effective electric field $\boldsymbol{E}_\mu = \partial_t\boldsymbol{A}_\mu - \nabla_{\mathbf{R}_\mu}\epsilon$, we condense Eq. (9) into the following compact form,

$$\frac{dT_{n,marg}}{dt} = \mathrm{Re}\langle\chi|\sum_{\mu=1}^{N_n}\boldsymbol{E}_\mu\cdot\hat{\boldsymbol{v}}_\mu|\chi\rangle_{\underline{\underline{\mathbf{R}}}}. \tag{10}$$

This equation is valid when there are no external forces driving the system. In the presence of external potentials acting on the electrons, those potentials will enter the electronic equation of motion leading to a modified electronic conditional wave function $\Phi_{\underline{\underline{\mathbf{R}}}}(\underline{\underline{\mathbf{r}}}, t)$ and hence to modified functions $\boldsymbol{A}_\mu(\underline{\underline{\mathbf{R}}}, t)$ and $\epsilon(\underline{\underline{\mathbf{R}}}, t)$, but the form of Eq. (10) remains unchanged. On the other hand, external potentials acting on the nuclei will enter Eq. (10) directly through the replacements $\boldsymbol{A}_\mu(\underline{\underline{\mathbf{R}}}, t) \to \boldsymbol{A}_\mu(\underline{\underline{\mathbf{R}}}, t) + \boldsymbol{A}_{ext,\mu}(\underline{\underline{\mathbf{R}}}, t)$ and $\epsilon(\underline{\underline{\mathbf{R}}}, t) \to \epsilon(\underline{\underline{\mathbf{R}}}, t) + v_{ext}(\underline{\underline{\mathbf{R}}}, t)$.

Eq. (10) casts the rate of change, $dT_{n,marg}/dt$, in the form of the classical work done per unit time by an electric field on a charged particle. We can obtain the rate of change of the full nuclear kinetic energy, $dT_n/dt$, by adding $dT_{n,geo}/dt$ to $dT_{n,marg}/dt$. A separate identity has been derived for $dT_{n,geo}/dt$ [31], but it cannot be expressed in a force-times-velocity form like Eq. (10).

In the absence of external forces, the total energy of the complete electron-nuclear system is conserved and can be written as the sum of three gauge-invariant terms: $E_{tot} = \langle \Psi(t)|\hat{H}|\Psi(t)\rangle_{\underline{\underline{\mathbf{r}}}\underline{\underline{\mathbf{R}}}} = T_{n,marg}(t) + T_{n,geo}(t) + E_{\rm BO}(t)$, where $E_{\rm BO}(t) = \langle \Psi(t)|\hat{T}_e + \hat{V}_{ee} + \hat{V}_{en} + \hat{V}_{nn}|\Psi(t)\rangle_{\underline{\underline{\mathbf{r}}}\underline{\underline{\mathbf{R}}}}$ . As we will demonstrate in the later sections of this Letter, nonadiabatic dynamical processes can be efficiently analyzed in terms of the energy transfer between these three gauge-invariant quantities (even when external forces are present). It will turn out that $T_{n,geo}$ is sizable, when a nonadiabatic population transfer between BO surfaces occurs. On the other hand, in the extreme adiabatic limit, when the factorization of the total wave function reduces to $\Psi(\underline{\underline{\mathbf{r}}}, \underline{\underline{\mathbf{R}}}, t) = \chi(\underline{\underline{\mathbf{R}}}, t)\Phi_{\rm BO}(\underline{\underline{\mathbf{r}}}|\underline{\underline{\mathbf{R}}})$, i.e. to a vibrational wave packet $\chi(\underline{\underline{\mathbf{R}}}, t)$ oscillating in a single BO surface, then $T_{n,geo}$ tends to be small, and one mainly observes the periodic energy transfer between $T_{n,marg}(t)$ and $E_{\rm BO}(t)$, i.e. the transfer between kinetic and potential energy of the oscillating wave packet. The expectation value $\langle \chi(t)|\hat{H}_n|\chi(t)\rangle_{\underline{\underline{\mathbf{R}}}}$ of the nuclear subsystem Hamiltonian is not gauge-invariant. However, interestingly, in the particular gauge defined by $\langle \Phi_{\underline{\underline{\mathbf{R}}}}(\underline{\underline{\mathbf{r}}}, t)|i\partial_t|\Phi_{\underline{\underline{\mathbf{R}}}}(\underline{\underline{\mathbf{r}}}, t)\rangle_{\underline{\underline{\mathbf{r}}}} = 0$ (which is the natural gauge when the electronic factor is identical with a static BO state), $\langle \chi(t)|\hat{H}_n|\chi(t)\rangle_{\underline{\underline{\mathbf{R}}}}$ becomes identical with $E_{tot}$, suggesting that the periodic exchange between kinetic and potential energy of the marginal subsystem can be interpreted as periodic energy transfer between electrons and nuclei.

Next, we derive the SEI for the nuclear momentum. Here instead of summing up the momenta of all nuclei, let us consider each individual $\boldsymbol{P}_\mu = \langle \chi|\hat{\boldsymbol{p}}_\mu|\chi\rangle_{\underline{\underline{\mathbf{R}}}}$, where $\hat{\boldsymbol{p}}_\mu = -i\nabla_{\mathbf{R}_\mu} + \boldsymbol{A}_\mu$. Once again, we use the Heisenberg equation of motion to arrive at

$$\begin{aligned}\frac{d\boldsymbol{P}_\mu}{dt} &= i\langle \chi|[\hat{H}_n, \hat{\boldsymbol{p}}_\mu]|\chi\rangle_{\underline{\underline{\mathbf{R}}}} + \langle \chi|\partial_t \hat{\boldsymbol{p}}_\mu|\chi\rangle_{\underline{\underline{\mathbf{R}}}} \\ &= \langle \chi|\boldsymbol{E}_\mu|\chi\rangle_{\underline{\underline{\mathbf{R}}}} + i\langle \chi|[\hat{t}_n, \hat{\boldsymbol{p}}_\mu]|\chi\rangle_{\underline{\underline{\mathbf{R}}}}. \qquad (11)\end{aligned}$$

Let us denote $\boldsymbol{Q}_\mu = i\langle \chi|[\hat{t}_n, \hat{\boldsymbol{p}}_\mu]|\chi\rangle_{\underline{\underline{\mathbf{R}}}}$. The fact that $d\boldsymbol{P}_\mu/dt$ and $\langle \chi|\boldsymbol{E}_\mu|\chi\rangle_{\underline{\underline{\mathbf{R}}}}$ are both real implies $\boldsymbol{Q}_\mu$ is real. By some algebra, we derive the $G$ ($G = X, Y, Z$) component of $\boldsymbol{Q}_\mu$ as given by

$$\begin{aligned}Q_\mu^G &= \mathrm{Re}\langle \chi| \sum_{\nu G'} (\partial_{G'_\nu} A_{G_\mu} - \partial_{G_\mu} A_{G'_\nu}) \hat{v}_{G'_\nu}|\chi\rangle_{\underline{\underline{\mathbf{R}}}} \\ &= \mathrm{Re}\langle \chi| \sum_{\nu G'} \mathcal{C}_{\nu\mu}^{G'G} \hat{v}_{G'_\nu}|\chi\rangle_{\underline{\underline{\mathbf{R}}}}. \qquad (12)\end{aligned}$$

Here $\mathcal{C}_{\nu\mu}^{G'G} \equiv \partial_{G'_\nu} A_{G_\mu} - \partial_{G_\mu} A_{G'_\nu}$ is the Berry curvature. By grouping $Q_\mu^G$ into intranuclear and internuclear contributions,

$$Q_\mu^G = \mathrm{Re}\langle \chi| \sum_{G'} (\mathcal{C}_{\mu\mu}^{G'G} \hat{v}_{G'_\mu} + \sum_{\nu \neq \mu} \mathcal{C}_{\nu\mu}^{G'G} \hat{v}_{G'_\nu})|\chi\rangle_{\underline{\underline{\mathbf{R}}}}, \qquad (13)$$

we can identify some classical analogs. In particular, the intranuclear curvature $\mathcal{C}_{\mu\mu}^{G'G}$ behaves like a classical magnetic field $\boldsymbol{B}_\mu$, where $\mathcal{C}_{\mu\mu}^{XY} = B_\mu^Z$, $\mathcal{C}_{\mu\mu}^{YZ} = B_\mu^X$ and $\mathcal{C}_{\mu\mu}^{ZX} = B_\mu^Y$. Upon summing over $G'$, these intranuclear terms lead to the following simple expression:

$$\sum_{G'} \hat{C}_{\mu\mu}^{G'G} \hat{v}_{G'_\mu} = (\boldsymbol{B}_\mu \times \hat{\boldsymbol{v}}_\mu)_G. \qquad (14)$$

The classical counterpart of Eq. (14) is the magnetic force acting on nucleus $\mu$, which combined with $\boldsymbol{E}_\mu$ in Eq. (11) leads to the generalized Lorentz force operator

$$\hat{\boldsymbol{F}}_\mu = \boldsymbol{E}_\mu + \boldsymbol{B}_\mu \times \hat{\boldsymbol{v}}_\mu. \qquad (15)$$

In contrast to the classical picture, here this force is an operator and the electromagnetic field is nucleus specific. Moreover, the appearance of the magnetic force as $\boldsymbol{B}_\mu \times \hat{\boldsymbol{v}}_\mu$ rather than $-\boldsymbol{B}_\mu \times \hat{\boldsymbol{v}}_\mu$ occurs because the sign in our definition of $\boldsymbol{A}_\mu$ is opposite of that in the conventional definition [32], which flips the sign of $\boldsymbol{A}_\mu$ in the definition of $\hat{\boldsymbol{p}}_\mu$.

On the other hand, the summation over the internuclear curvature terms has no classical analog and does not readily simplify. Instead, we introduce an internuclear magnetic force operator

$$\hat{D}_\mu^G = \sum_{\nu \neq \mu, G'} \mathcal{C}_{\nu\mu}^{G'G} \hat{v}_{G'_\nu} = \sum_{\nu \neq \mu} (\nabla_{\mathbf{R}_\nu} A_{G_\mu} - \partial_{G_\mu} \boldsymbol{A}_\nu) \cdot \hat{\boldsymbol{v}}_\nu, \qquad (16)$$

The internuclear Berry curvature $\mathcal{C}_{\nu\mu}^{G'G}$ has been studied previously in the Born-Oppenheimer approximation [33, 34]. Eq. (11) simplifies to the following SEI:

$$\frac{d\boldsymbol{P}_\mu}{dt} = \mathrm{Re}\langle \chi|\hat{\boldsymbol{F}}_\mu + \hat{\boldsymbol{D}}_\mu|\chi\rangle_{\underline{\underline{\mathbf{R}}}} \equiv \mathrm{Re}\langle \chi|\hat{\boldsymbol{\mathcal{F}}}_\mu|\chi\rangle_{\underline{\underline{\mathbf{R}}}}. \qquad (17)$$

Here $\hat{\boldsymbol{\mathcal{F}}}_\mu \equiv \hat{\boldsymbol{F}}_\mu + \hat{\boldsymbol{D}}_\mu$ denotes the electromagnetic-like *force operator* whose expectation value gives the mean force on nucleus $\mu$. Our $\hat{\boldsymbol{\mathcal{F}}}_\mu$ has a formal resemblance to the classical force function $\boldsymbol{F}^I_\mu$ that was introduced in previous works for calculating the time evolution of the nuclear momentum $\boldsymbol{P}^I_\mu$ of a particular trajectory $\underline{\underline{\mathbf{R}}}^I(t)$ in a trajectory-based representation of the nuclear Schrödinger equation [35–40]. In fact, one can show that $\boldsymbol{P}^I_\mu \equiv \boldsymbol{P}_\mu(\underline{\underline{\mathbf{R}}}^I(t),t) = \mathrm{Re}\frac{\hat{\boldsymbol{p}}_\mu\chi}{\chi}\Big|_{\underline{\underline{\mathbf{R}}}=\underline{\underline{\mathbf{R}}}^I(t)}$ and $\boldsymbol{F}^I_\mu \equiv \boldsymbol{F}_\mu(\underline{\underline{\mathbf{R}}}^I(t),t) = \mathrm{Re}\frac{\hat{\boldsymbol{\mathcal{F}}}_\mu\chi}{\chi}\Big|_{\underline{\underline{\mathbf{R}}}=\underline{\underline{\mathbf{R}}}^I(t)}$. Although $\boldsymbol{P}^I_\mu$ and $\boldsymbol{F}^I_\mu$ are auxiliary quantities tied to the trajectory based methods, where $\underline{\underline{\mathbf{R}}}$ and $t$ are no longer independent variables, one expects to recover $d\boldsymbol{P}_\mu/dt$ upon taking the ensemble average. In this Letter we have derived a representation independent identity for the rate of change of the observable $\langle\chi|\hat{\boldsymbol{p}}_\mu|\chi\rangle$, showing that it is governed by the novel force operator on the RHS of Eq. (17). Interestingly, the RHS can be evaluated in the position representation by replacing $\hat{\boldsymbol{\mathcal{F}}}_\mu$ by $\boldsymbol{F}_\mu(\underline{\underline{\mathbf{R}}},t)$ due to the formal resemblance of these forces, see [41] and example below.

Next, we show that the same force operator appears in the equations of motion for $\boldsymbol{L}_\mu$ and $T_{n,marg}$. By following a similar derivation, we can connect the rate of change of angular momentum with an effective torque, [41]

$$\frac{d\boldsymbol{L}_\mu}{dt} = \mathrm{Re}\langle\chi|\mathbf{R}_\mu \times \hat{\boldsymbol{\mathcal{F}}}_\mu|\chi\rangle_{\underline{\underline{\mathbf{R}}}}. \qquad (18)$$

On the contrary, such a simple relation does not hold for kinetic energy of an individual nucleus, i.e., $dT_{\mu,marg}/dt \neq \mathrm{Re}\langle\chi|\hat{\boldsymbol{\mathcal{F}}}_\mu\cdot\hat{\boldsymbol{v}}_\mu|\chi\rangle_{\underline{\underline{\mathbf{R}}}}$. Replacing $\hat{\boldsymbol{\mathcal{F}}}_\mu$ by $\hat{\boldsymbol{F}}_\mu$ or $\boldsymbol{E}_\mu$ does not lead to the right formula either. Only by summing over all nuclei can we achieve an equality involving $\hat{\boldsymbol{\mathcal{F}}}_\mu$, [41]

$$\frac{dT_{n,marg}}{dt} = \mathrm{Re}\langle\chi|\sum_{\mu=1}^{N_n}\hat{\boldsymbol{\mathcal{F}}}_\mu\cdot\hat{\boldsymbol{v}}_\mu|\chi\rangle_{\underline{\underline{\mathbf{R}}}}. \qquad (19)$$

Eq. (10) and Eq. (19) imply that the magnetic forces do no work. One can immediately see this for the intranuclear magnetic force because $(\boldsymbol{B}_\mu \times \hat{\boldsymbol{v}}_\mu)\cdot\hat{\boldsymbol{v}}_\mu = 0$. On the other hand, although each $\mathrm{Re}\langle\chi|\hat{\boldsymbol{D}}_\mu\cdot\hat{\boldsymbol{v}}_\mu|\chi\rangle$ is nonzero, they compensate one another upon summation due to the internal nature of the force $\hat{\boldsymbol{D}}_\mu$.

To visualize the SEIs, we design an exactly solvable model of two nuclei moving in one dimension. Let us assume that the information of the electron-nuclear wave function $\Psi$ has been condensed into a nuclear wave function $\chi$ that satisfies the following TDSE:

$$i\partial_t\chi = \frac{1}{2M}\sum_{\mu=1}^{2}(-i\partial_{X_\mu}+A_\mu)^2\chi + \epsilon\chi. \qquad (20)$$

Although $\Psi$ is a complex function, we can always choose a gauge such that $\chi$ is real. Here instead of following the conventional way of solving for $\chi$ with given time dependent scalar and vector potentials and initial condition, we go the other way around. By choosing a particular form of time-evolving wave function $\chi(X_1,X_2,t)$, we aim to find analytical forms of the corresponding $A_1, A_2$ and $\epsilon$ as functions of $X_1, X_2$ and $t$ that yield such a $\chi$. In this work, we choose $\chi$ to be a normalized Gaussian function of a fixed width $\sigma$,

$$\chi(X_1,X_2,t) = \frac{1}{\sigma\sqrt{\pi}}\exp\Big\{-\frac{1}{2\sigma^2}\sum_{\mu=1}^{2}[X_\mu - g_\mu(t)]^2\Big\}, \qquad (21)$$

whose center moves along a trajectory $(g_1(t), g_2(t))$. Taking the real part of Eq. (20), and using the fact that $\chi$ is chosen to be real, we can deduce the form of $\epsilon$ in terms of $\chi$ and $A_\mu$. Here we choose the following vector potentials yielding $\chi$:

$$A_1(X_1,X_2,t) = f(t)[-X_2 + g_2(t)] + Mg_1'(t), \qquad (22)$$

$$A_2(X_1,X_2,t) = f(t)[X_1 - g_1(t)] + Mg_2'(t), \qquad (23)$$

where $f(t)$ and $g_\mu$ are periodic functions in time, yielding a spiral nuclear center trajectory (see [41] for details).

Because the nuclei move in one dimension, the intranuclear magnetic field $B$ is absent so that the generalized Lorentz force reduces to the electromotive force, $F_\mu = E_\mu$. However, propagating the Newton's equation with only this force $\bar{F}_\mu = \langle\chi|F_\mu|\chi\rangle$ will lead to an incorrect trajectory for the center of the nuclear wave packet. Due to the presence of a nonzero internuclear Berry curvature $\mathcal{C}_{12} = \partial_{X_1}A_2 - \partial_{X_2}A_1 = 2f(t)$, the internuclear magnetic force operators $\hat{D}_1 = -\mathcal{C}_{12}\hat{v}_2$ and $\hat{D}_2 = \mathcal{C}_{12}\hat{v}_1$ contribute to the correct force $\bar{F}_\mu + \bar{D}_\mu = \mathrm{Re}\langle\chi|F_\mu + \hat{D}_\mu|\chi\rangle$ acting on the nuclear wave packet.

This is illustrated in Fig 1 (a), where we compare trajectories generated by the exact and incomplete mean force. As can be seen, the exact force drives the nuclear center to swirl around its initial position, while the incomplete force produces a spurious acceleration of the second nucleus in the $X$ direction. Although the missing $\hat{D}_\mu$ contributions are small relative to $\hat{F}_\mu$ (see [41]), its cumulative effect in time can be sizable. After 10 cycles, this leads to a net displacement of $X_2$ of about 0.1 Bohr. In molecular dynamics, the size of $\hat{\boldsymbol{D}}_\mu$ and its relative importance are unknown and deserve further study.

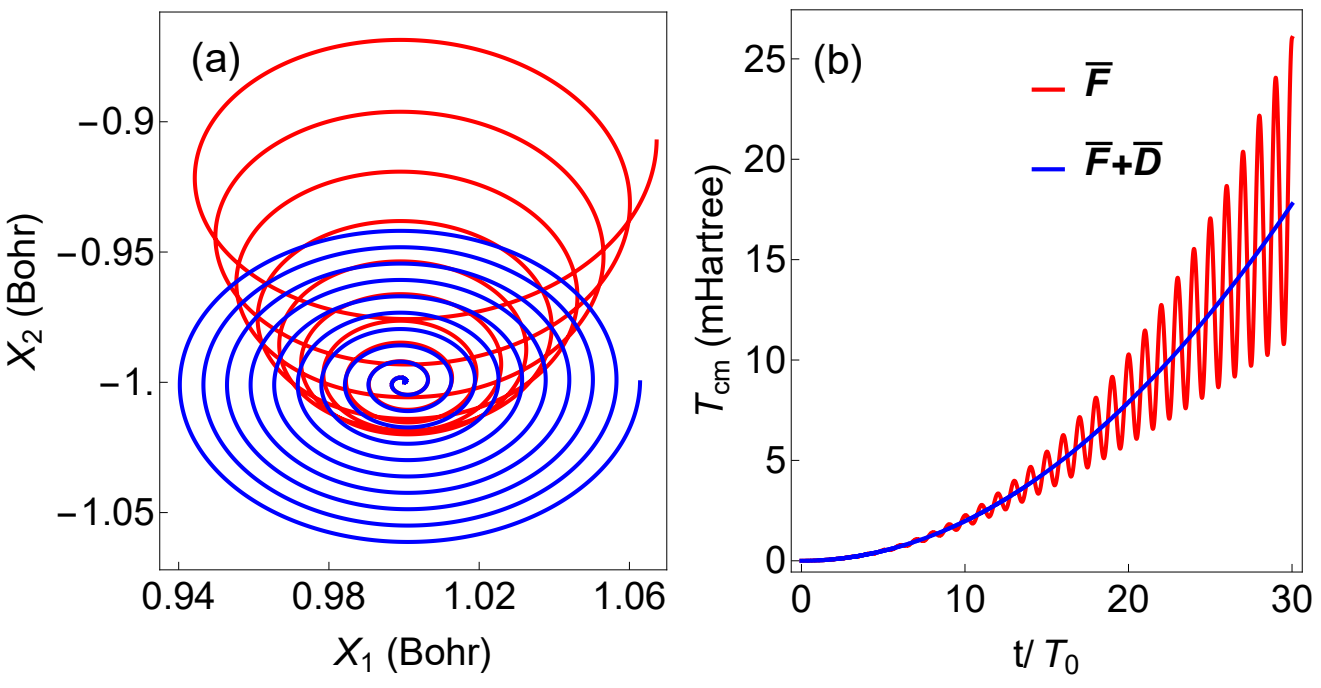

FIG. 1. (a) Trajectory of the center of the nuclear wavepacket shown as a 2D plot. The initial position is at $(1, -1)$. With the incomplete force $\bar{\boldsymbol{F}}$, the nuclear center deviates from the correct spiral trajectory after a few cycles. (b) Center of mass kinetic energy $T_{\rm cm}$ as a function of $t$ calculated with $\bar{\boldsymbol{F}}$ and $\bar{\boldsymbol{F}} + \bar{\boldsymbol{D}}$. Here $T_0 = 6.8$ fs is the time period.

Next, let us consider the kinetic energy change during this swirling process. By direct computation, the kinetic energy increase over the initial time is

$$T_{n,marg}(t) - T_{n,marg}(0) = \frac{1}{2M}\sigma^2 f^2(t) + \frac{1}{2}M\bar{\boldsymbol{v}}^2. \quad (24)$$

Here the first term on the RHS of Eq. (24) is purely a quantum effect. If we assume that the nuclear wave packet is narrow and $f(t)$ is small, this term is only of secondary importance. By contrast, the dominant term is the second one, where $\bar{\boldsymbol{v}} = \Big(g_1'(t), g_2'(t)\Big)$ is the velocity of the nuclear center of mass and $\frac{1}{2}M\bar{\boldsymbol{v}}^2 \equiv T_{\rm cm}$ is the classical nuclear kinetic energy. In Fig 1 (b), we compare $T_{\rm cm}$ using the trajectories generated by the exact and incomplete forces. As shown, the incomplete force yields $T_{\rm cm}$ with artificial oscillations around the exact curve.

Our model has great similarity with the current-driven atomic water wheel model studied in [9], if we interpret $X_1$ and $X_2$ as two cartesian coordinates of the same nucleus. We emphasize that $\bar{\boldsymbol{F}} + \bar{\boldsymbol{D}}$ is the key quantity that determines the stability and working efficiency of the water wheel, see [41] for further discussions.

In Fig 2, we also simulate the energy transfer between electrons and nuclei after a vertical excitation to an excited electronic state in a two-level model diatomic molecule (details can be found in the supplemental material [41]). This is related to the early stage of a bond-breaking process [42], particularly the reaction pathways of small molecules [43], or the redistribution of energy of DNA and RNA [3, 4]. Here we idealize the two BO surfaces as harmonic potentials with an avoided crossing in the region $R = 0.5 - 1$ Bohr. As illustrated in Fig 2(a), after vertical excitation from the equilibrium position of the lower surface, the nuclear wave packet $\chi$ starts propagating on the upper one. If we neglect the coupling of the two surfaces, $\chi$ undergoes perfect harmonic oscillations around $R = 0.5$ Bohr with a time period $T_0 = 13.6$ fs. However, because of the coupling of the surfaces, the nuclear wave packet splits into two branches as it goes through the avoided crossing. One branch stays on the upper surface while the other transfers to the lower one accompanied by energy transfer from electrons to nuclear kinetic energy. This process also breaks the time periodicity because the split wave packets feel the gradients (forces) of two different BO potentials. This is also reflected in Fig 2(c), where the rate of change of $T_{n,marg}$ and $T_n$ as functions of $t$ deviate from periodic behavior when $t > \frac{T_0}{2}$. Here we focus on two points, $t = \frac{T_0}{8}$ and $t = \frac{5T_0}{8}$, roughly corresponding to the first two maxima. As shown, there is a significant decrease in $dT_{n,marg}/dt$ from $t = \frac{T_0}{8}$ to $t = \frac{5T_0}{8}$. To understand this change, we use our identity (19) and rewrite it as

$$\frac{dT_{n,marg}}{dt} = \int |\chi(R,t)|^2 F(R,t) v(R,t) dR, \quad (25)$$

where we introduce a velocity function $v(R,t) = \mathrm{Re}\frac{\hat{v}\chi}{\chi}$. In Fig 2(b) and (d) we compare $|\chi|^2$, $F$ and $v$ at the chosen times. As shown, in (d) $|\chi|^2$ is broadened; and in the region where $|\chi|^2$ is large, the exact $F$ and $v$ become smaller in absolute value although they change sign as a result of the reversal of the moving direction. All of these changes contribute to the drop of $dT_{n,marg}/dt$ at $t = \frac{5T}{8}$.

As a side remark, the relative importance of $dT_{n,geo}/dt$ can be deduced from Fig 2(c). It is unimportant for $t < \frac{5T_0}{8}$ but starts to build up when $\chi$ repeatedly traverses the nonadiabatic region. The example shown here corresponds to an intermediate nonadiabatic coupling regime. In the adiabatic regime, the effect of $T_{n,geo}$ becomes much smaller. Discussions on this along with the electronic energy and population transfer in adiabatic regimes are presented in the supplemental material [41].

To better understand the exact force function, we compare it with the forces calculated from the BO surfaces, which are straight lines in Fig 2(b) and (d). As shown in (b), the exact force has similar slope with $F_{\rm BO2}$, although amplified by quantum corrections. In (d), however, it follows the slope of $F_{\rm BO1}$ for $R < 1$ Bohr while turning to follow $F_{\rm BO2}$ for $R > 1$ Bohr, suggesting a split of the nuclear wave packet at $R = 1$ Bohr, as also indicated by the emerging shoulder in $|\chi|^2$. Our observation of the piecewise behavior in the exact force in case of wave packet splitting is also in line with the literature. [30]

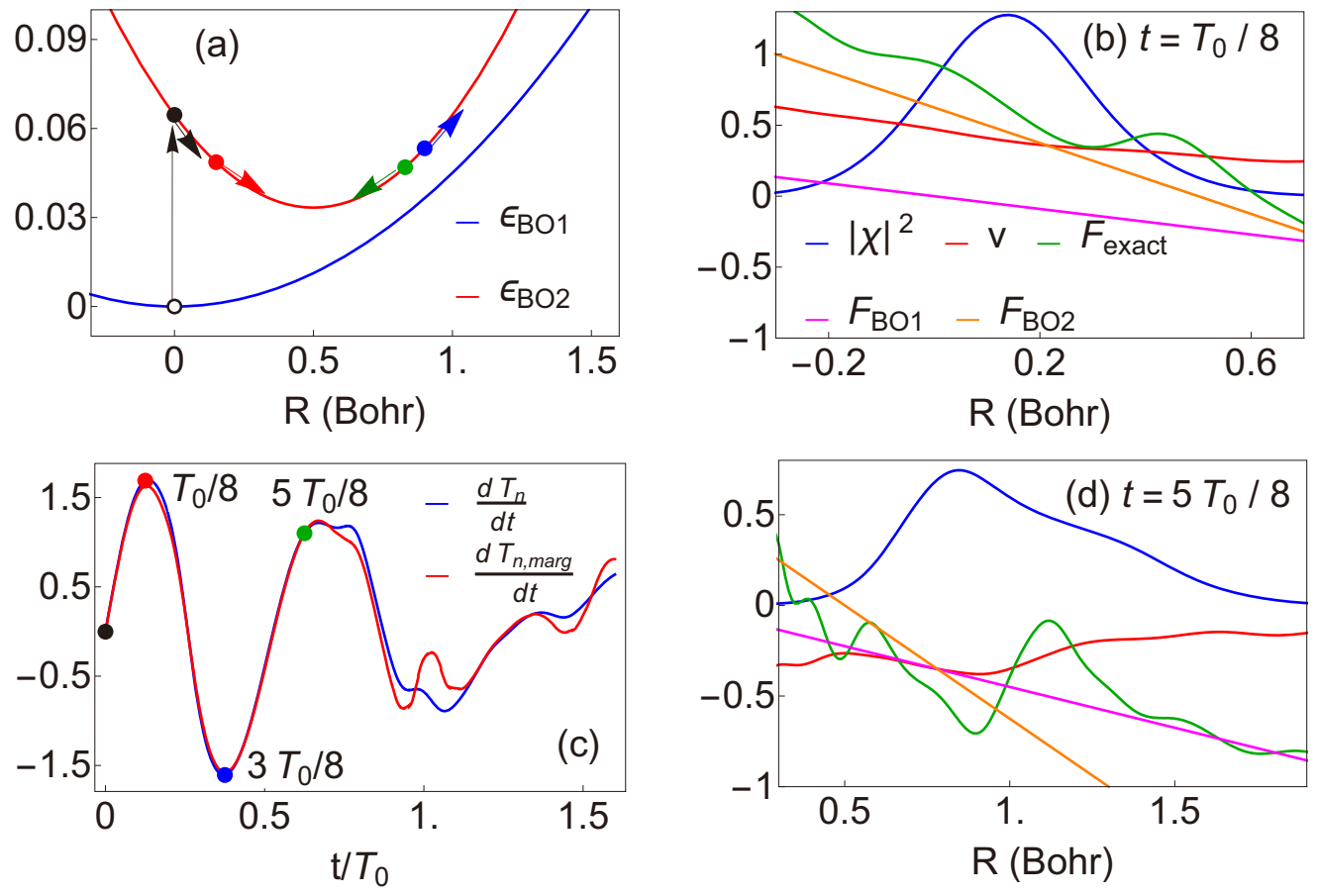

FIG. 2. (a) BO surfaces (in Hartree) with illustrative dots and arrows indicating the nuclear vibration after a vertical excitation. (b) $|\chi|^2$, $v = \mathrm{Re}\frac{\hat{v}\chi}{\chi}$ and the exact force function in atomic units evaluated at $t = \frac{T_0}{8}$ where $T_0 = 13.6$ fs. Forces from the static BO surfaces are also shown as reference curves. To put all these curves in one figure, $|\chi|^2$ has been scaled by a factor of $\frac{1}{2}$, $v$ by a factor of 100 and all the forces by a factor of 5. (c) rate of change of $T_n$ and $T_{n,marg}$ as functions of time in unit of $10^{-2}$Hartree/fs. Chosen instants of time are marked by colored dots corresponding to the ones in (a). (d) same as in (b) except evaluated at $t = \frac{5T_0}{8}$.

To summarize, in this Letter we have used the nuclear TDSE of the exact factorization method to establish subsystem Ehrenfest identities for the three main canonical variables in classical and quantum mechanics. We have shown that the same effective electromagnetic force operator $\hat{\boldsymbol{\mathcal{F}}}_\mu$ appears in all three identities. The magnetic component of the corresponding electromagnetic field comes from two sources: (a) the more familiar intranuclear Berry curvature associated with different cartesian coordinates of the same nucleus [32]; (b) the internuclear Berry curvature calculated with arbitrary cartesian coordinates of two different nuclei. (a) has the classical interpretation of an effective magnetic field acting on a given nucleus, while (b) has no classical analog. In practical calculations, one can take advantage of the recently developed exact factorization-based density functional theory [44–47] to evaluate these forces as functionals of the conditional electronic density and nuclear probability density. One can also apply the approximate Born-Oppenheimer factorization to derive the same SEIs for approximate forces. By condensing the enormous amount of information in the electron-nuclear wave function into comprehensible quantities such as force and velocity, we will gain more insight into dynamical processes on the atomic scale.

We thank Ali Abedi and Federica Agostini for helpful discussions. This project has received funding from the European Research Council (ERC) under the European Unions Horizon 2020 research and innovation programme (grant agreement No. ERC-2017-AdG-788890).

---

* chenlichem@pku.edu.cn

Supporting Information of

# Energy, Momentum and Angular Momentum Transfer Between Electrons and Nuclei

Chen Li[1,2,3], Ryan Requist[1,2] and E. K. U. Gross[1,2]

[1] *Max Planck Institute of Microstructure Physics, Weinberg 2, 06120, Halle, Germany*

[2] *Fritz Haber Center for Molecular Dynamics, Institute of Chemistry, The Hebrew University of Jerusalem, Jerusalem 91904 Israel*

[3] *Beijing National Laboratory for Molecular Sciences, College of Chemistry and Molecular Engineering, Peking University, Beijing 100871, China*

**Contents**

## I. SOME DETAILS ON DERIVING THE SUBSYSTEM EHRENFEST IDENTITIES (SEIS)

### A. Calculation of $\boldsymbol{Q}_\mu$

In the main text, we have introduced $\boldsymbol{Q}_\mu$ as

$$\begin{aligned}\boldsymbol{Q}_\mu &= i\langle\chi|[\hat{t}_n,\hat{\boldsymbol{p}}_\mu]|\chi\rangle_{\underline{\underline{\mathbf{R}}}} = \sum_\nu i\langle\chi|[(-i\nabla_{\mathbf{R}_\nu}+\boldsymbol{A}_\nu)^2,(-i\nabla_{\mathbf{R}_\mu}+\boldsymbol{A}_\mu)]|\chi\rangle_{\underline{\underline{\mathbf{R}}}} \\ &= \sum_\nu \frac{1}{2M_\nu}\Big(\langle\chi|[(-i\nabla_{\mathbf{R}_\nu}+\boldsymbol{A}_\nu)^2,\nabla_{\mathbf{R}_\mu}]|\chi\rangle_{\underline{\underline{\mathbf{R}}}} + i\langle\chi|[(-i\nabla_{\mathbf{R}_\nu}+\boldsymbol{A}_\nu)^2,\boldsymbol{A}_\mu]|\chi\rangle_{\underline{\underline{\mathbf{R}}}}\Big) \\ &= \sum_\nu \frac{1}{2M_\nu}(\boldsymbol{Q}^1_{\nu\mu}+\boldsymbol{Q}^2_{\nu\mu}). \end{aligned} \tag{S1}$$

Here

$$\boldsymbol{Q}^1_{\nu\mu} = \langle\chi|[(-i\nabla_{\mathbf{R}_\nu}+\boldsymbol{A}_\nu)^2,\nabla_{\mathbf{R}_\mu}]|\chi\rangle_{\underline{\underline{\mathbf{R}}}}, \tag{S2}$$

$$\boldsymbol{Q}^2_{\nu\mu} = i\langle\chi|[(-i\nabla_{\mathbf{R}_\nu}+\boldsymbol{A}_\nu)^2,\boldsymbol{A}_\mu]|\chi\rangle_{\underline{\underline{\mathbf{R}}}}. \tag{S3}$$

In the following derivations, we will repeatedly use the following two identities to simplify commutators,

$$[\nabla, O]\chi = (\nabla O)\chi, \tag{S4}$$

$$[\nabla^2, O]\chi = (\nabla^2 O)\chi + 2(\nabla O)\cdot\nabla\chi. \tag{S5}$$

Here $O$ is an arbitrary non-differential operator and $\nabla$ can be arbitrary gradient. Additionally, we have the following identity for the kinetic energy operator,

$$(-i\nabla_{\mathbf{R}_\nu}+\boldsymbol{A}_\nu)^2 = -\nabla^2_{\mathbf{R}_\nu} - 2i\boldsymbol{A}_\nu\cdot\nabla_{\mathbf{R}_\nu} - i(\nabla_{\mathbf{R}_\nu}\cdot\boldsymbol{A}_\nu) + \boldsymbol{A}^2_\nu, \tag{S6}$$

Using the above identities, let us first calculate $\boldsymbol{Q}^1_{\nu\mu}$ as follows,

$$\begin{aligned}\boldsymbol{Q}^1_{\nu\mu} &= \langle\chi|[-\nabla^2_{\mathbf{R}_\nu} - 2i\boldsymbol{A}_\nu\cdot\nabla_{\mathbf{R}_\nu} - i(\nabla_{\mathbf{R}_\nu}\cdot\boldsymbol{A}_\nu) + \boldsymbol{A}^2_\nu, \nabla_{\mathbf{R}_\mu}]|\chi\rangle_{\underline{\underline{\mathbf{R}}}} \\ &= \langle\chi|[-2i\boldsymbol{A}_\nu\cdot\nabla_{\mathbf{R}_\nu},\nabla_{\mathbf{R}_\mu}]|\chi\rangle_{\underline{\underline{\mathbf{R}}}} + \langle\chi|[-i(\nabla_{\mathbf{R}_\nu}\cdot\boldsymbol{A}_\nu)+\boldsymbol{A}^2_\nu,\nabla_{\mathbf{R}_\mu}]|\chi\rangle_{\underline{\underline{\mathbf{R}}}} \\ &= \langle\chi|[-2i\boldsymbol{A}_\nu\cdot\nabla_{\mathbf{R}_\nu},\nabla_{\mathbf{R}_\mu}]|\chi\rangle_{\underline{\underline{\mathbf{R}}}} - \langle\chi|\Big(\nabla_{\mathbf{R}_\mu}(-i\nabla_{\mathbf{R}_\nu}\cdot\boldsymbol{A}_\nu+\boldsymbol{A}^2_\nu)\Big)|\chi\rangle_{\underline{\underline{\mathbf{R}}}}. \end{aligned} \tag{S7}$$

In the last equality, we have used identity (S4). Next we calculate $\boldsymbol{Q}^2_{\nu\mu}$.

$$\begin{aligned}\boldsymbol{Q}^2_{\nu\mu} &= i\langle\chi|[-\nabla^2_{\mathbf{R}_\nu} - 2i\boldsymbol{A}_\nu\cdot\nabla_{\mathbf{R}_\nu} - i(\nabla_{\mathbf{R}_\nu}\cdot\boldsymbol{A}_\nu) + \boldsymbol{A}^2_\nu, \boldsymbol{A}_\mu]|\chi\rangle_{\underline{\underline{\mathbf{R}}}} \\ &= i\langle\chi|[-\nabla^2_{\mathbf{R}_\nu} - 2i\boldsymbol{A}_\nu\cdot\nabla_{\mathbf{R}_\nu},\boldsymbol{A}_\mu]|\chi\rangle_{\underline{\underline{\mathbf{R}}}} \\ &= -i\langle\chi|[\nabla^2_{\mathbf{R}_\nu},\boldsymbol{A}_\mu]|\chi\rangle_{\underline{\underline{\mathbf{R}}}} + 2\langle\chi|(\boldsymbol{A}_\nu\cdot\nabla_{\mathbf{R}_\nu})\boldsymbol{A}_\mu|\chi\rangle_{\underline{\underline{\mathbf{R}}}}. \end{aligned} \tag{S8}$$

Then taking the $G(G=X,Y,Z)$ component of Eq. (S7) and Eq. (S8) leads to

$$\begin{aligned}\boldsymbol{Q}_{\nu\mu}^{1G} &= \langle\chi|[-2i\boldsymbol{A}_\nu\cdot\nabla_{\mathbf{R}_\nu},\partial_{G_\mu}]|\chi\rangle_{\underline{\underline{\mathbf{R}}}} - \langle\chi|\Big(\partial_{G_\mu}(-i\nabla_{\mathbf{R}_\nu}\cdot\boldsymbol{A}_\nu+\boldsymbol{A}_\nu^2)\Big)|\chi\rangle_{\underline{\underline{\mathbf{R}}}} \\ &= 2i\langle\chi|(\partial_{G_\mu}\boldsymbol{A}_\nu)\cdot\nabla_{\mathbf{R}_\nu}|\chi\rangle_{\underline{\underline{\mathbf{R}}}} - 2\langle\chi|\Big((\partial_{G_\mu}\boldsymbol{A}_\nu)\cdot\boldsymbol{A}_\nu\Big)|\chi\rangle_{\underline{\underline{\mathbf{R}}}} + i\langle\chi|\Big(\partial_{G_\mu}(\nabla_{\mathbf{R}_\nu}\cdot\boldsymbol{A}_\nu)\Big)|\chi\rangle_{\underline{\underline{\mathbf{R}}}} \\ &= -2\langle\chi|(\partial_{G_\mu}\boldsymbol{A}_\nu)\cdot(-i\nabla_{\mathbf{R}_\nu}+\boldsymbol{A}_\nu)|\chi\rangle_{\underline{\underline{\mathbf{R}}}} + i\langle\chi|\Big(\partial_{G_\mu}(\nabla_{\mathbf{R}_\nu}\cdot\boldsymbol{A}_\nu)\Big)|\chi\rangle_{\underline{\underline{\mathbf{R}}}}, \end{aligned} \tag{S9}$$

and

$$\begin{aligned}\boldsymbol{Q}_{\nu\mu}^{2G} &= -i\langle\chi|[\nabla^2_{\mathbf{R}_\nu},A_{G_\mu}]|\chi\rangle_{\underline{\underline{\mathbf{R}}}} + 2\langle\chi|(\boldsymbol{A}_\nu\cdot\nabla_{\mathbf{R}_\nu})A_{G_\mu}|\chi\rangle_{\underline{\underline{\mathbf{R}}}} \\ &= -i\langle\chi|(\nabla^2_{\mathbf{R}_\nu}A_{G_\mu})|\chi\rangle_{\underline{\underline{\mathbf{R}}}} - 2i\langle\chi|(\nabla_{\mathbf{R}_\nu}A_{G_\mu})\cdot\nabla_{\mathbf{R}_\nu}|\chi\rangle_{\underline{\underline{\mathbf{R}}}} + 2\langle\chi|\boldsymbol{A}_\nu\cdot(\nabla_{\mathbf{R}_\nu}A_{G_\mu})|\chi\rangle_{\underline{\underline{\mathbf{R}}}} \\ &= 2\langle\chi|(\nabla_{\mathbf{R}_\nu}A_{G_\mu})\cdot(-i\nabla_{\mathbf{R}_\nu}+\boldsymbol{A}_\nu)|\chi\rangle_{\underline{\underline{\mathbf{R}}}} - i\langle\chi|(\nabla^2_{\mathbf{R}_\nu}A_{G_\mu})|\chi\rangle_{\underline{\underline{\mathbf{R}}}}. \end{aligned} \tag{S10}$$

Now let us add up Eqs. (S9) and Eq. (S10), multiply by $\frac{1}{2M_\nu}$ and sum over $\nu$. We obtain the $G$ component of $\boldsymbol{Q}_\mu$ the following,

$$Q_\mu^G = \sum_\nu \langle\chi|(\nabla_{\mathbf{R}_\nu}A_{G_\mu}-\partial_{G_\mu}\boldsymbol{A}_\nu)\cdot\hat{\boldsymbol{v}}_\nu|\chi\rangle_{\underline{\underline{\mathbf{R}}}} + i\sum_\nu \frac{1}{2M_\nu}\langle\chi|\Big(\partial_{G_\mu}(\nabla_{\mathbf{R}_\nu}\cdot\boldsymbol{A}_\nu)-\nabla^2_{\mathbf{R}_\nu}A_{G_\mu}\Big)|\chi\rangle_{\underline{\underline{\mathbf{R}}}}. \tag{S11}$$

Here $\hat{\boldsymbol{v}}_\nu = \frac{1}{M_\nu}(-i\nabla_{\mathbf{R}_\nu}+\boldsymbol{A}_\nu)$. The second term in the RHS of Eq. (S11) is purely imaginary while the LHS is real. Therefore, the second term on the RHS must compensate the imaginary part of the first term. And what is left is the real part of Eq. (S11),

$$\begin{aligned}Q_\mu^G &= \mathrm{Re}\sum_\nu\langle\chi|(\nabla_{\mathbf{R}_\nu}A_{G_\mu}-\partial_{G_\mu}\boldsymbol{A}_\nu)\cdot\hat{\boldsymbol{v}}_\nu|\chi\rangle_{\underline{\underline{\mathbf{R}}}} \\ &= \mathrm{Re}\langle\chi|\sum_{\nu G'}(\partial_{G'_\nu}A_{G_\mu}-\partial_{G_\mu}A_{G'_\nu})\hat{v}_{G'_\nu}|\chi\rangle_{\underline{\underline{\mathbf{R}}}}, \qquad (G=X,Y,Z). \end{aligned} \tag{S12}$$

### B. Proving the SEI for nuclear angular momentum

The angular momentum of nucleus $\mu$ is defined as the expectation value $\boldsymbol{L}_\mu = \langle\chi|\hat{\boldsymbol{L}}_\mu|\chi\rangle_{\underline{\underline{\mathbf{R}}}}$, where $\hat{\boldsymbol{L}}_\mu = \mathbf{R}_\mu\times(-i\nabla_{\mathbf{R}_\mu}+\boldsymbol{A}_\mu)$. The Heisenberg equation of motion for $\boldsymbol{L}_\mu$ reads

$$\begin{aligned}\frac{d\boldsymbol{L}_\mu}{dt} &= i\langle\chi|[\hat{H}_n,\hat{\boldsymbol{L}}_\mu]|\chi\rangle_{\underline{\underline{\mathbf{R}}}} + \langle\chi|\partial_t\hat{\boldsymbol{L}}_\mu|\chi\rangle_{\underline{\underline{\mathbf{R}}}} \\ &= i\langle\chi|\Big[\sum_{\nu=1}^{N_n}\frac{1}{2M_\nu}(-i\nabla_{\mathbf{R}_\nu}+\boldsymbol{A}_\nu)^2+\epsilon,\hat{\boldsymbol{L}}_\mu\Big]|\chi\rangle_{\underline{\underline{\mathbf{R}}}} + \langle\chi|\partial_t\hat{\boldsymbol{L}}_\mu|\chi\rangle_{\underline{\underline{\mathbf{R}}}} \\ &= i\langle\chi|\Big[\sum_{\nu=1}^{N_n}\frac{1}{2M_\nu}(-i\nabla_{\mathbf{R}_\nu}+\boldsymbol{A}_\nu)^2,\hat{\boldsymbol{L}}_\mu\Big]|\chi\rangle_{\underline{\underline{\mathbf{R}}}} + i\langle\chi|[\epsilon,\hat{\boldsymbol{L}}_\mu]|\chi\rangle_{\underline{\underline{\mathbf{R}}}} + \langle\chi|\partial_t\hat{\boldsymbol{L}}_\mu|\chi\rangle_{\underline{\underline{\mathbf{R}}}} \\ &= \boldsymbol{L}_\mu^1+\boldsymbol{L}_\mu^2. \end{aligned} \tag{S13}$$

Here

$$\boldsymbol{L}_\mu^1 = i\langle\chi|\Big[\sum_{\nu=1}^{N_n}\frac{1}{2M_\nu}(-i\nabla_{\mathbf{R}_\nu}+\boldsymbol{A}_\nu)^2, \hat{\boldsymbol{L}}_\mu\Big]|\chi\rangle_{\underline{\underline{\mathbf{R}}}}, \tag{S14}$$

$$\boldsymbol{L}_\mu^2 = i\langle\chi|[\epsilon, \hat{\boldsymbol{L}}_\mu]|\chi\rangle_{\underline{\underline{\mathbf{R}}}} + \langle\chi|\partial_t\hat{\boldsymbol{L}}_\mu|\chi\rangle_{\underline{\underline{\mathbf{R}}}}. \tag{S15}$$

We first calculate $\boldsymbol{L}_\mu^2$.

$$\begin{aligned}
\boldsymbol{L}_\mu^2 &= i\langle\chi|[\epsilon, \mathbf{R}_\mu\times(-i\nabla_{\mathbf{R}_\mu}+\boldsymbol{A}_\mu)]|\chi\rangle_{\underline{\underline{\mathbf{R}}}} + \langle\chi|\partial_t\Big(\mathbf{R}_\mu\times(-i\nabla_{\mathbf{R}_\mu}+\boldsymbol{A}_\mu)\Big)|\chi\rangle_{\underline{\underline{\mathbf{R}}}}\\
&= \langle\chi|[\epsilon, \mathbf{R}_\mu\times\nabla_{\mathbf{R}_\mu}] + \mathbf{R}_\mu\times(\partial_t\boldsymbol{A}_\mu)|\chi\rangle_{\underline{\underline{\mathbf{R}}}}\\
&= \langle\chi|\mathbf{R}_\mu\times(\partial_t\boldsymbol{A}_\mu - \nabla_{\mathbf{R}_\mu}\epsilon)|\chi\rangle_{\underline{\underline{\mathbf{R}}}}\\
&= \langle\chi|\mathbf{R}_\mu\times\boldsymbol{E}_\mu|\chi\rangle_{\underline{\underline{\mathbf{R}}}}.
\end{aligned} \tag{S16}$$

Here $\boldsymbol{E}_\mu$ is the induced electromotive force (EMF) acting on nucleus $\mu$. Next we calculate $\boldsymbol{L}_\mu^1$. It suffices to calculate the following commutator,

$$\begin{aligned}
\boldsymbol{U}^{\nu\mu} &= i\langle\chi|\Big[(-i\nabla_{\mathbf{R}_\nu}+\boldsymbol{A}_\nu)^2, \hat{\boldsymbol{L}}_\mu\Big]|\chi\rangle_{\underline{\underline{\mathbf{R}}}}\\
&= i\langle\chi|\Big[(-i\nabla_{\mathbf{R}_\nu}+\boldsymbol{A}_\nu)^2, \mathbf{R}_\mu\times(-i\nabla_{\mathbf{R}_\mu}+\boldsymbol{A}_\mu)\Big]|\chi\rangle_{\underline{\underline{\mathbf{R}}}}\\
&= \boldsymbol{U}_1^{\nu\mu} + \boldsymbol{U}_2^{\nu\mu}.
\end{aligned} \tag{S17}$$

Here,

$$\begin{aligned}
\boldsymbol{U}_1^{\nu\mu} &= \langle\chi|\Big[(-i\nabla_{\mathbf{R}_\nu}+\boldsymbol{A}_\nu)^2, \mathbf{R}_\mu\times\nabla_{\mathbf{R}_\mu}\Big]|\chi\rangle_{\underline{\underline{\mathbf{R}}}}\\
\boldsymbol{U}_2^{\nu\mu} &= \langle\chi|\Big[(-i\nabla_{\mathbf{R}_\nu}+\boldsymbol{A}_\nu)^2, \mathbf{R}_\mu\times\boldsymbol{A}_\mu\Big]|\chi\rangle_{\underline{\underline{\mathbf{R}}}}.
\end{aligned} \tag{S18}$$

Let us first consider the $Z$ component of $\boldsymbol{U}_1^{\nu\mu}$.

$$\begin{aligned}
U_{1Z}^{\nu\mu} &= \langle\chi|\Big[(-i\nabla_{\mathbf{R}_\nu}+\boldsymbol{A}_\nu)^2, (X_\mu\partial_{Y_\mu} - Y_\mu\partial_{X_\mu})\Big]|\chi\rangle_{\underline{\underline{\mathbf{R}}}}\\
&= \langle\chi|\Big[-\nabla_{\mathbf{R}_\nu}^2 - 2i\boldsymbol{A}_\nu\cdot\nabla_{\mathbf{R}_\nu} - i(\nabla_{\mathbf{R}_\nu}\cdot\boldsymbol{A}_\nu) + \boldsymbol{A}_\nu^2, (X_\mu\partial_{Y_\mu} - Y_\mu\partial_{X_\mu})\Big]|\chi\rangle_{\underline{\underline{\mathbf{R}}}}\\
&= U_{1Z1}^{\nu\mu} + U_{1Z2}^{\nu\mu} + U_{1Z3}^{\nu\mu}.
\end{aligned} \tag{S19}$$

Here $U_{1Zj}^{\nu\mu}$ ($j = 1, 2, 3$) refer to the expectation value of (i) $-\nabla_{\mathbf{R}_\nu}^2$, (ii) $-2i\boldsymbol{A}_\nu\cdot\nabla_{\mathbf{R}_\nu}$ and (iii) $-i(\nabla_{\mathbf{R}_\nu}\cdot\boldsymbol{A}_\nu) + \boldsymbol{A}_\nu^2$ taking commutator with $X_\mu\partial_{Y_\mu} - Y_\mu\partial_{X_\mu}$, which we evaluate one by one.
(i) $U_{1Z1}^{\nu\mu}$.

$$U_{1Z1}^{\nu\mu} = \langle\chi|\Big[-\nabla_{\mathbf{R}_\nu}^2, (X_\mu\partial_{Y_\mu} - Y_\mu\partial_{X_\mu})\Big]|\chi\rangle_{\underline{\underline{\mathbf{R}}}} = \langle\chi|\Big[-\nabla_{\mathbf{R}_\mu}^2, (X_\mu\partial_{Y_\mu} - Y_\mu\partial_{X_\mu})\Big]|\chi\rangle_{\underline{\underline{\mathbf{R}}}}\delta_{\mu\nu}. \tag{S20}$$

Here

$$\langle\chi|[-\nabla^2_{\mathbf{R}_\mu}, X_\mu\partial_{Y_\mu}]|\chi\rangle_{\underline{\underline{\mathbf{R}}}} = -\langle\chi|\nabla^2_{\mathbf{R}_\mu}(X_\mu\partial_{Y_\mu}\chi) - X_\mu\partial_{Y_\mu}(\nabla^2_{\mathbf{R}_\mu}\chi)\rangle_{\underline{\underline{\mathbf{R}}}}$$
$$= -\langle\chi|2(\nabla_{\mathbf{R}_\mu}X_\mu)\cdot\nabla_{\mathbf{R}_\mu}(\partial_{Y_\mu}\chi)\rangle_{\underline{\underline{\mathbf{R}}}}$$
$$= -\langle\chi|2\partial_{X_\mu}(\partial_{Y_\mu}\chi)\rangle_{\underline{\underline{\mathbf{R}}}}. \tag{S21}$$

In the second equality we have used $\nabla^2_{\mathbf{R}_\mu}X_\mu = 0$. Similarly, we can show that

$$\langle\chi|[-\nabla^2_{\mathbf{R}_\mu}, Y_\mu\partial_{X_\mu}]|\chi\rangle_{\underline{\underline{\mathbf{R}}}} = -\langle\chi|2\partial_{Y_\mu}(\partial_{X_\mu}\chi)\rangle_{\underline{\underline{\mathbf{R}}}}. \tag{S22}$$

Therefore,

$$\langle\chi|[-\nabla^2_{\mathbf{R}_\mu}, X_\mu\partial_{Y_\mu} - Y_\mu\partial_{X_\mu}]|\chi\rangle_{\underline{\underline{\mathbf{R}}}} = -\langle\chi|2\partial_{X_\mu}(\partial_{Y_\mu}\chi) - 2\partial_{Y_\mu}(\partial_{X_\mu}\chi)\rangle_{\underline{\underline{\mathbf{R}}}} = 0, \tag{S23}$$

and hence $U^{\nu\mu}_{1Z1} = 0$.

(ii) $U^{\nu\mu}_{1Z2}$.

$$U^{\nu\mu}_{1Z2} = \langle\chi|\Big[-2i\boldsymbol{A}_\nu\cdot\nabla_{\mathbf{R}_\nu}, (X_\mu\partial_{Y_\mu} - Y_\mu\partial_{X_\mu})\Big]|\chi\rangle_{\underline{\underline{\mathbf{R}}}}. \tag{S24}$$

Here,

$$\langle\chi|\Big[-2i\boldsymbol{A}_\nu\cdot\nabla_{\mathbf{R}_\nu}, X_\mu\partial_{Y_\mu}\Big]|\chi\rangle_{\underline{\underline{\mathbf{R}}}} = -2i\langle\chi|\boldsymbol{A}_\nu\cdot\nabla_{\mathbf{R}_\nu}(X_\mu\partial_{Y_\mu}\chi) - X_\mu\partial_{Y_\mu}(\boldsymbol{A}_\nu\cdot\nabla_{\mathbf{R}_\nu}\chi)\rangle_{\underline{\underline{\mathbf{R}}}}$$
$$= -2i\langle\chi|(\partial_{Y_\mu}\chi)\boldsymbol{A}_\nu\cdot\nabla_{\mathbf{R}_\nu}X_\mu - X_\mu(\partial_{Y_\mu}\boldsymbol{A}_\nu)\cdot\nabla_{\mathbf{R}_\nu}\chi\rangle_{\underline{\underline{\mathbf{R}}}}$$
$$= -2i\langle\chi|(\partial_{Y_\mu}\chi)A_{X_\mu}\delta_{\mu\nu} - X_\mu(\partial_{Y_\mu}\boldsymbol{A}_\nu)\cdot\nabla_{\mathbf{R}_\nu}\chi\rangle_{\underline{\underline{\mathbf{R}}}}. \tag{S25}$$

Interchanging $X$ and $Y$ leads to

$$\langle\chi|\Big[-2i\boldsymbol{A}_\nu\cdot\nabla_{\mathbf{R}_\nu}, Y_\mu\partial_{X_\mu}\Big]|\chi\rangle_{\underline{\underline{\mathbf{R}}}} = -2i\langle\chi|(\partial_{X_\mu}\chi)A_{Y_\mu}\delta_{\mu\nu} - Y_\mu(\partial_{X_\mu}\boldsymbol{A}_\nu)\cdot\nabla_{\mathbf{R}_\nu}\chi\rangle_{\underline{\underline{\mathbf{R}}}}. \tag{S26}$$

Therefore,

$$U^{\nu\mu}_{1Z2} = \langle\chi|\Big[-2i\boldsymbol{A}_\nu\cdot\nabla_{\mathbf{R}_\nu}, (X_\mu\partial_{Y_\mu} - Y_\mu\partial_{X_\mu})\Big]|\chi\rangle_{\underline{\underline{\mathbf{R}}}}$$
$$= \langle\chi| - 2i\Big((\partial_{Y_\mu}\chi)A_{X_\mu} - (\partial_{X_\mu}\chi)A_{Y_\mu}\Big)\delta_{\mu\nu} + 2i\Big(X_\mu(\partial_{Y_\mu}\boldsymbol{A}_\nu) - Y_\mu(\partial_{X_\mu}\boldsymbol{A}_\nu)\Big)\cdot\nabla_{\mathbf{R}_\nu}\chi\rangle_{\underline{\underline{\mathbf{R}}}}$$
$$= \langle\chi| - 2i\Big(\boldsymbol{A}_\mu\times\nabla_{\mathbf{R}_\mu}\chi\Big)_Z\delta_{\mu\nu} + 2i\Big((X_\mu\partial_{Y_\mu} - Y_\mu\partial_{X_\mu})\boldsymbol{A}_\nu)\Big)\cdot\nabla_{\mathbf{R}_\nu}\chi\rangle_{\underline{\underline{\mathbf{R}}}}. \tag{S27}$$

(iii) $U^{\nu\mu}_{1Z3}$.

$$U^{\nu\mu}_{1Z3} = \langle\chi|\Big[-i(\nabla_{\mathbf{R}_\nu}\cdot\boldsymbol{A}_\nu) + \boldsymbol{A}^2_\nu, (X_\mu\partial_{Y_\mu} - Y_\mu\partial_{X_\mu})\Big]|\chi\rangle_{\underline{\underline{\mathbf{R}}}}$$
$$= -\langle\chi|(X_\mu\partial_{Y_\mu} - Y_\mu\partial_{X_\mu})\Big(-i(\nabla_{\mathbf{R}_\nu}\cdot\boldsymbol{A}_\nu) + \boldsymbol{A}^2_\nu\Big)|\chi\rangle_{\underline{\underline{\mathbf{R}}}}$$
$$= i\langle\chi|\Big((\mathbf{R}_\mu\times\nabla_{\mathbf{R}_\mu})_Z(\nabla_{\mathbf{R}_\nu}\cdot\boldsymbol{A}_\nu)\Big)|\chi\rangle_{\underline{\underline{\mathbf{R}}}} - 2\langle\chi|\boldsymbol{A}_\nu\cdot\Big((X_\mu\partial_{Y_\mu} - Y_\mu\partial_{X_\mu})\boldsymbol{A}_\nu\Big)|\chi\rangle_{\underline{\underline{\mathbf{R}}}}. \tag{S28}$$

Next, we consider the $Z$ component of $\boldsymbol{U}_2^{\nu\mu}$.

$$\begin{aligned}
U_{2Z}^{\nu\mu} &= i\langle\chi|\Big[(-i\nabla_{\mathbf{R}_\nu}+\boldsymbol{A}_\nu)^2,(X_\mu A_{Y_\mu}-Y_\mu A_{X_\mu})\Big]|\chi\rangle_{\underline{\underline{\mathbf{R}}}}\\
&= i\langle\chi|\Big[-\nabla^2_{\mathbf{R}_\nu}-2i\boldsymbol{A}_\nu\cdot\nabla_{\mathbf{R}_\nu}-i(\nabla_{\mathbf{R}_\nu}\cdot\boldsymbol{A}_\nu)+\boldsymbol{A}_\nu^2,(X_\mu A_{Y_\mu}-Y_\mu A_{X_\mu})\Big]|\chi\rangle_{\underline{\underline{\mathbf{R}}}}\\
&= i\langle\chi|\Big[-\nabla^2_{\mathbf{R}_\nu},(X_\mu A_{Y_\mu}-Y_\mu A_{X_\mu})\Big]|\chi\rangle_{\underline{\underline{\mathbf{R}}}}+\langle\chi|\Big[2\boldsymbol{A}_\nu\cdot\nabla_{\mathbf{R}_\nu},(X_\mu A_{Y_\mu}-Y_\mu A_{X_\mu})\Big]|\chi\rangle_{\underline{\underline{\mathbf{R}}}}\\
&= U_{2Z1}^{\nu\mu}+U_{2Z2}^{\nu\mu}.
\end{aligned}\tag{S29}$$

Here $U_{2Z1}^{\nu\mu}$ and $U_{2Z2}^{\nu\mu}$ denote the two terms on the RHS, respectively.

(i) $U_{2Z1}^{\nu\mu}$.

$$U_{2Z1}^{\nu\mu} = i\langle\chi|\Big[-\nabla^2_{\mathbf{R}_\nu},(X_\mu A_{Y_\mu}-Y_\mu A_{X_\mu})\Big]|\chi\rangle_{\underline{\underline{\mathbf{R}}}}.\tag{S30}$$

Here,

$$\begin{aligned}
i\langle\chi|[-\nabla^2_{\mathbf{R}_\nu},X_\mu A_{Y_\mu}]|\chi\rangle_{\underline{\underline{\mathbf{R}}}} &= -i\langle\chi|\Big(\nabla^2_{\mathbf{R}_\nu}(X_\mu A_{Y_\mu})\Big)|\chi\rangle_{\underline{\underline{\mathbf{R}}}}-2i\langle\chi|\Big(\nabla_{\mathbf{R}_\nu}(X_\mu A_{Y_\mu})\Big)\cdot\nabla_{\mathbf{R}_\nu}|\chi\rangle_{\underline{\underline{\mathbf{R}}}}\\
&= -i\langle\chi|(2\nabla_{\mathbf{R}_\nu}X_\mu\cdot\nabla_{\mathbf{R}_\nu}A_{Y_\mu}+X_\mu\nabla^2_{\mathbf{R}_\nu}A_{Y_\mu})|\chi\rangle_{\underline{\underline{\mathbf{R}}}}\\
&\quad -2i\langle\chi|(A_{Y_\mu}\nabla_{\mathbf{R}_\nu}X_\mu+X_\mu\nabla_{\mathbf{R}_\nu}A_{Y_\mu})\cdot\nabla_{\mathbf{R}_\nu}|\chi\rangle_{\underline{\underline{\mathbf{R}}}}\\
&= -i\langle\chi|(2\delta_{\mu\nu}\partial_{X_\mu}A_{Y_\mu}+X_\mu\nabla^2_{\mathbf{R}_\nu}A_{Y_\mu})|\chi\rangle_{\underline{\underline{\mathbf{R}}}}\\
&\quad -2i\langle\chi|A_{Y_\mu}\delta_{\mu\nu}\partial_{X_\mu}+X_\mu(\nabla_{\mathbf{R}_\nu}A_{Y_\mu})\cdot\nabla_{\mathbf{R}_\nu}|\chi\rangle_{\underline{\underline{\mathbf{R}}}}.
\end{aligned}\tag{S31}$$

Interchanging $X$ and $Y$ leads to

$$\begin{aligned}
i\langle\chi|[-\nabla^2_{\mathbf{R}_\nu},Y_\mu A_{X_\mu}]|\chi\rangle_{\underline{\underline{\mathbf{R}}}} &= -i\langle\chi|(2\delta_{\mu\nu}\partial_{Y_\mu}A_{X_\mu}+Y_\mu\nabla^2_{\mathbf{R}_\nu}A_{X_\mu})|\chi\rangle_{\underline{\underline{\mathbf{R}}}}\\
&\quad -2i\langle\chi|A_{X_\mu}\delta_{\mu\nu}\partial_{Y_\mu}+Y_\mu(\nabla_{\mathbf{R}_\nu}A_{X_\mu})\cdot\nabla_{\mathbf{R}_\nu}|\chi\rangle_{\underline{\underline{\mathbf{R}}}}.
\end{aligned}\tag{S32}$$

Therefore,

$$\begin{aligned}
U_{2Z1}^{\nu\mu} &= i\langle\chi|\Big[-\nabla^2_{\mathbf{R}_\nu},(X_\mu A_{Y_\mu}-Y_\mu A_{X_\mu})\Big]|\chi\rangle_{\underline{\underline{\mathbf{R}}}}\\
&= -i\langle\chi|(2\delta_{\mu\nu}\partial_{X_\mu}A_{Y_\mu}+X_\mu\nabla^2_{\mathbf{R}_\nu}A_{Y_\mu})-(2\delta_{\mu\nu}\partial_{Y_\mu}A_{X_\mu}+Y_\mu\nabla^2_{\mathbf{R}_\nu}A_{X_\mu})|\chi\rangle_{\underline{\underline{\mathbf{R}}}}\\
&\quad -2i\langle\chi|A_{Y_\mu}\delta_{\mu\nu}\partial_{X_\mu}+X_\mu(\nabla_{\mathbf{R}_\nu}A_{Y_\mu})\cdot\nabla_{\mathbf{R}_\nu}-A_{X_\mu}\delta_{\mu\nu}\partial_{Y_\mu}-Y_\mu(\nabla_{\mathbf{R}_\nu}A_{X_\mu})\cdot\nabla_{\mathbf{R}_\nu}|\chi\rangle_{\underline{\underline{\mathbf{R}}}}\\
&= -i\langle\chi|2\delta_{\mu\nu}(\nabla_{\mathbf{R}_\mu}\times\boldsymbol{A}_\mu)_Z+(X_\mu\nabla^2_{\mathbf{R}_\nu}A_{Y_\mu}-Y_\mu\nabla^2_{\mathbf{R}_\nu}A_{X_\mu})|\chi\rangle_{\underline{\underline{\mathbf{R}}}}\\
&\quad -2i\langle\chi|-\delta_{\mu\nu}(\boldsymbol{A}_\mu\times\nabla_{\mathbf{R}_\mu})_Z+\Big(X_\mu(\nabla_{\mathbf{R}_\nu}A_{Y_\mu})-Y_\mu(\nabla_{\mathbf{R}_\nu}A_{X_\mu})\Big)\cdot\nabla_{\mathbf{R}_\nu}|\chi\rangle_{\underline{\underline{\mathbf{R}}}}.
\end{aligned}\tag{S33}$$

(ii) $U_{2Z2}^{\nu\mu}$.

$$U_{2Z2}^{\nu\mu} = \langle\chi|\Big[2\boldsymbol{A}_\nu\cdot\nabla_{\mathbf{R}_\nu},(X_\mu A_{Y_\mu}-Y_\mu A_{X_\mu})\Big]|\chi\rangle_{\underline{\underline{\mathbf{R}}}}.\tag{S34}$$

Here,

$$\begin{aligned}\langle\chi|[2\boldsymbol{A}_\nu\cdot\nabla_{\mathbf{R}_\nu}, X_\mu A_{Y_\mu}]|\chi\rangle_{\underline{\underline{\mathbf{R}}}} &= 2\langle\chi|\boldsymbol{A}_\nu\cdot\Big(\nabla_{\mathbf{R}_\nu}(X_\mu A_{Y_\mu})\Big)|\chi\rangle_{\underline{\underline{\mathbf{R}}}}\\ &= 2\langle\chi|\boldsymbol{A}_\nu\cdot\Big(A_{Y_\mu}\nabla_{\mathbf{R}_\nu}X_\mu + X_\mu\nabla_{\mathbf{R}_\nu}A_{Y_\mu})\Big)|\chi\rangle_{\underline{\underline{\mathbf{R}}}}\\ &= 2\langle\chi|\delta_{\mu\nu}A_{X_\mu}A_{Y_\mu} + X_\mu\boldsymbol{A}_\nu\cdot(\nabla_{\mathbf{R}_\nu}A_{Y_\mu})|\chi\rangle_{\underline{\underline{\mathbf{R}}}}.\end{aligned}\tag{S35}$$

Interchanging $X$ and $Y$, we have

$$\langle\chi|[2\boldsymbol{A}_\nu\cdot\nabla_{\mathbf{R}_\nu}, Y_\mu A_{X_\mu}]|\chi\rangle_{\underline{\underline{\mathbf{R}}}} = 2\langle\chi|\delta_{\mu\nu}A_{Y_\mu}A_{X_\mu} + Y_\mu\boldsymbol{A}_\nu\cdot(\nabla_{\mathbf{R}_\nu}A_{X_\mu})|\chi\rangle_{\underline{\underline{\mathbf{R}}}}.\tag{S36}$$

Therefore,

$$\begin{aligned}U^{\nu\mu}_{2Z2} &= \langle\chi|\Big[2\boldsymbol{A}_\nu\cdot\nabla_{\mathbf{R}_\nu},(X_\mu A_{Y_\mu}-Y_\mu A_{X_\mu})\Big]|\chi\rangle_{\underline{\underline{\mathbf{R}}}}\\ &= 2\langle\chi|\boldsymbol{A}_\nu\cdot(X_\mu\nabla_{\mathbf{R}_\nu}A_{Y_\mu} - Y_\mu\nabla_{\mathbf{R}_\nu}A_{X_\mu})|\chi\rangle_{\underline{\underline{\mathbf{R}}}}.\end{aligned}\tag{S37}$$

Now let us add up all the ingredients in $U^{\nu\mu}_Z$. Since only the real part of $U^{\nu\mu}_Z$ has contribution to $\frac{d\boldsymbol{L}_\mu}{dt}$, it suffices to collect the real part of Eq. (S27)+Eq. (S28)+ Eq. (S33) + Eq. (S37),

$$\begin{aligned}\mathrm{Re}\,U^{\nu\mu}_Z &= \mathrm{Re}\Big(U^{\nu\mu}_{1Z1}+U^{\nu\mu}_{1Z2}+U^{\nu\mu}_{1Z3}+U^{\nu\mu}_{2Z1}+U^{\nu\mu}_{2Z2}\Big)\\ &= \mathrm{Re}\Big\{\langle\chi| - 2i\Big(\boldsymbol{A}_\mu\times\nabla_{\mathbf{R}_\mu}\chi\Big)_Z\delta_{\mu\nu} + 2i\Big((X_\mu\partial_{Y_\mu}-Y_\mu\partial_{X_\mu})\boldsymbol{A}_\nu\Big)\cdot\nabla_{\mathbf{R}_\nu}\chi\rangle_{\underline{\underline{\mathbf{R}}}}\\ &\quad - 2\langle\chi|\boldsymbol{A}_\nu\cdot\Big((X_\mu\partial_{Y_\mu}-Y_\mu\partial_{X_\mu})\boldsymbol{A}_\nu\Big)|\chi\rangle_{\underline{\underline{\mathbf{R}}}}\\ &\quad + \langle\chi|2i\delta_{\mu\nu}(\boldsymbol{A}_\mu\times\nabla_{\mathbf{R}_\mu})_Z - 2i\Big(X_\mu(\nabla_{\mathbf{R}_\nu}A_{Y_\mu}) - Y_\mu(\nabla_{\mathbf{R}_\nu}A_{X_\mu})\Big)\cdot\nabla_{\mathbf{R}_\nu}|\chi\rangle_{\underline{\underline{\mathbf{R}}}}\\ &\quad + 2\langle\chi|\boldsymbol{A}_\nu\cdot(X_\mu\nabla_{\mathbf{R}_\nu}A_{Y_\mu} - Y_\mu\nabla_{\mathbf{R}_\nu}A_{X_\mu})|\chi\rangle_{\underline{\underline{\mathbf{R}}}}\Big\}.\end{aligned}\tag{S38}$$

After cancelation and rearranging the terms, we have

$$\begin{aligned}\mathrm{Re}\,U^{\nu\mu}_Z &= \mathrm{Re}\Big\{-2\langle\chi|\Big(X_\mu(\partial_{Y_\mu}\boldsymbol{A}_\nu) - Y_\mu(\partial_{X_\mu}\boldsymbol{A}_\nu)\Big)\cdot(-i\nabla_{\mathbf{R}_\nu}+\boldsymbol{A}_\nu)|\chi\rangle_{\underline{\underline{\mathbf{R}}}}\\ &\quad + 2\langle\chi|\Big(X_\mu(\nabla_{\mathbf{R}_\nu}A_{Y_\mu}) - Y_\mu(\nabla_{\mathbf{R}_\nu}A_{X_\mu})\Big)\cdot(-i\nabla_{\mathbf{R}_\nu}+\boldsymbol{A}_\nu)|\chi\rangle_{\underline{\underline{\mathbf{R}}}}\\ &= 2M_\nu\mathrm{Re}\langle\chi|\Big(X_\mu(\nabla_{\mathbf{R}_\nu}A_{Y_\mu}-\partial_{Y_\mu}\boldsymbol{A}_\nu) - Y_\mu(\nabla_{\mathbf{R}_\nu}A_{X_\mu}-\partial_{X_\mu}\boldsymbol{A}_\nu)\Big)\cdot\hat{\boldsymbol{v}}_\nu|\chi\rangle_{\underline{\underline{\mathbf{R}}}}.\end{aligned}\tag{S39}$$

When $\mu=\nu$,

$$\begin{aligned}X_\mu(\nabla_{\mathbf{R}_\mu}A_{Y_\mu}-\partial_{Y_\mu}\boldsymbol{A}_\mu)\cdot\hat{\boldsymbol{v}}_\mu &= X_\mu\Big[(\partial_{X_\mu}A_{Y_\mu}-\partial_{Y_\mu}A_{X_\mu})\hat{v}_{X_\mu} + (\partial_{Z_\mu}A_{Y_\mu}-\partial_{Y_\mu}A_{Z_\mu})\hat{v}_{Z_\mu}\Big]\\ &= X_\mu(B_{Z_\mu}\hat{v}_{X_\mu} - B_{X_\mu}\hat{v}_{Z_\mu})\\ &= X_\mu(\boldsymbol{B}_\mu\times\hat{\boldsymbol{v}}_\mu)_Y.\end{aligned}\tag{S40}$$

Here we have used the definition of the induced nuclear magnetic field, $B_{X_\mu} = \partial_{Y_\mu} A_{Z_\mu} - \partial_{Z_\mu} A_{Y_\mu}, B_{Y_\mu} = \partial_{Z_\mu} A_{X_\mu} - \partial_{X_\mu} A_{Z_\mu}$ and $B_{Z_\mu} = \partial_{X_\mu} A_{Y_\mu} - \partial_{Y_\mu} A_{X_\mu}$. Interchanging $X$ and $Y$ leads to

$$Y_\mu(\nabla_{\mathbf{R}_\mu} A_{X_\mu} - \partial_{X_\mu} \boldsymbol{A}_\mu) \cdot \hat{\boldsymbol{v}}_\mu = Y_\mu(\boldsymbol{B}_\mu \times \hat{\boldsymbol{v}}_\mu)_X. \tag{S41}$$

Therefore, one can achieve the following simplification to the operator in Eq. (S39),

$$\begin{aligned}
&\Big(X_\mu(\nabla_{\mathbf{R}_\mu} A_{Y_\mu} - \partial_{Y_\mu} \boldsymbol{A}_\mu) - Y_\mu(\nabla_{\mathbf{R}_\mu} A_{X_\mu} - \partial_{X_\mu} \boldsymbol{A}_\mu)\Big) \cdot \hat{\boldsymbol{v}}_\mu \\
&= X_\mu(\boldsymbol{B}_\mu \times \hat{\boldsymbol{v}}_\mu)_Y - Y_\mu(\boldsymbol{B}_\mu \times \hat{\boldsymbol{v}}_\mu)_X \\
&= \Big(\boldsymbol{R}_\mu \times (\boldsymbol{B}_\mu \times \hat{\boldsymbol{v}}_\mu)\Big)_Z.
\end{aligned} \tag{S42}$$

Now taking the $Z$ component of Eq. (S14), we have

$$\begin{aligned}
L^1_{\mu Z} &= \text{Re } L^1_{\mu Z} = \sum_{\nu=1}^{N_n} \frac{1}{2M_\nu} \text{Re } U^{\nu\mu}_Z \\
&= \frac{1}{2M_\mu} \text{Re } U^{\mu\mu}_Z + \sum_{\nu \neq \mu} \frac{1}{2M_\nu} \text{Re} \langle \chi | \Big( X_\mu(\nabla_{\mathbf{R}_\nu} A_{Y_\mu} - \partial_{Y_\mu} \boldsymbol{A}_\nu) - Y_\mu(\nabla_{\mathbf{R}_\nu} A_{X_\mu} - \partial_{X_\mu} \boldsymbol{A}_\nu) \Big) \cdot \hat{\boldsymbol{v}}_\nu | \chi \rangle_{\underline{\underline{\mathbf{R}}}} \\
&= \text{Re} \langle \chi | \Big( \boldsymbol{R}_\mu \times (\boldsymbol{B}_\mu \times \hat{\boldsymbol{v}}_\mu) \Big)_Z | \chi \rangle_{\underline{\underline{\mathbf{R}}}} + \text{Re} \langle \chi | X_\mu \sum_{\nu \neq \mu} (\nabla_{\mathbf{R}_\nu} A_{Y_\mu} - \partial_{Y_\mu} \boldsymbol{A}_\nu) \cdot \hat{\boldsymbol{v}}_\nu | \chi \rangle_{\underline{\underline{\mathbf{R}}}} \\
&\quad - \text{Re} \langle \chi | Y_\mu \sum_{\nu \neq \mu} (\nabla_{\mathbf{R}_\nu} A_{X_\mu} - \partial_{X_\mu} \boldsymbol{A}_\nu) \cdot \hat{\boldsymbol{v}}_\nu | \chi \rangle_{\underline{\underline{\mathbf{R}}}} \\
&= \text{Re} \langle \chi | \Big( \boldsymbol{R}_\mu \times (\boldsymbol{B}_\mu \times \hat{\boldsymbol{v}}_\mu) \Big)_Z | \chi \rangle_{\underline{\underline{\mathbf{R}}}} + \text{Re} \langle \chi | X_\mu \hat{D}^Y_\mu - Y_\mu \hat{D}^X_\mu | \chi \rangle_{\underline{\underline{\mathbf{R}}}} \\
&= \text{Re} \langle \chi | \Big( \boldsymbol{R}_\mu \times (\boldsymbol{B}_\mu \times \hat{\boldsymbol{v}}_\mu + \hat{\boldsymbol{D}}_\mu) \Big)_Z | \chi \rangle_{\underline{\underline{\mathbf{R}}}}.
\end{aligned} \tag{S43}$$

Here we have used our definition of $\hat{D}^G_\mu$ in the main text as $\hat{D}^G_\mu = \sum_{\nu \neq \mu} (\nabla_{\mathbf{R}_\nu} A_{G_\mu} - \partial_{G_\mu} \boldsymbol{A}_\nu) \cdot \hat{\boldsymbol{v}}_\nu$. One can follow the same procedure and derive the similar equalities for the $X$ and $Y$ components of $\boldsymbol{L}^1_\mu$. Thus we conclude that

$$\boldsymbol{L}^1_\mu = \text{Re} \langle \chi | \boldsymbol{R}_\mu \times (\boldsymbol{B}_\mu \times \hat{\boldsymbol{v}}_\mu + \hat{\boldsymbol{D}}_\mu) | \chi \rangle_{\underline{\underline{\mathbf{R}}}}. \tag{S44}$$

Substituting Eq. (S44) and Eq. (S16) into Eq. (S13), we arrive at

$$\frac{d\boldsymbol{L}_\mu}{dt} = \text{Re} \langle \chi | \boldsymbol{R}_\mu \times \hat{\boldsymbol{\mathcal{F}}}_\mu | \chi \rangle_{\underline{\underline{\mathbf{R}}}}, \tag{S45}$$

where $\hat{\boldsymbol{\mathcal{F}}}_\mu = \hat{\boldsymbol{F}}_\mu + \hat{\boldsymbol{D}}_\mu = \boldsymbol{E}_\mu + \boldsymbol{B}_\mu \times \hat{\boldsymbol{v}}_\mu + \hat{\boldsymbol{D}}_\mu$. Here $\hat{\boldsymbol{F}}_\mu$ is the generalized Lorentz force.

### C. Proof that the internuclear magnetic force does no work

In the main text, we have shown that

$$\frac{dT_{n,marg}}{dt} = \mathrm{Re}\langle\chi|\sum_{\mu=1}^{N_n}\boldsymbol{E}_\mu\cdot\hat{\boldsymbol{v}}_\mu|\chi\rangle_{\underline{\underline{\mathbf{R}}}} = \mathrm{Re}\langle\chi|\sum_{\mu=1}^{N_n}\hat{\boldsymbol{F}}_\mu\cdot\hat{\boldsymbol{v}}_\mu|\chi\rangle_{\underline{\underline{\mathbf{R}}}}. \tag{S46}$$

Here the second equality is because the intranuclear magnetic force does no work.

Now we further prove that the internuclear magnetic force does no work either, i.e.,

$$\mathrm{Re}\langle\chi|\sum_{\mu=1}^{N_n}\hat{\boldsymbol{D}}_\mu\cdot\hat{\boldsymbol{v}}_\mu|\chi\rangle_{\underline{\underline{\mathbf{R}}}} = 0, \tag{S47}$$

so that

$$\frac{dT_{n,marg}}{dt} = \mathrm{Re}\langle\chi|\sum_{\mu=1}^{N_n}\hat{\boldsymbol{\mathcal{F}}}_\mu\cdot\hat{\boldsymbol{v}}_\mu|\chi\rangle_{\underline{\underline{\mathbf{R}}}}. \tag{S48}$$

To prove Eq. (S47), let us insert the expression of $\hat{\boldsymbol{D}}_\mu$ in it,

$$\begin{aligned}\mathrm{Re}\langle\chi|\sum_{\mu=1}^{N_n}\hat{\boldsymbol{D}}_\mu\cdot\hat{\boldsymbol{v}}_\mu|\chi\rangle_{\underline{\underline{\mathbf{R}}}} &= \mathrm{Re}\langle\chi|\sum_{\mu=1}^{N_n}\sum_{G}\hat{D}_\mu^G\hat{v}_{G_\mu}|\chi\rangle_{\underline{\underline{\mathbf{R}}}} \\ &= \mathrm{Re}\langle\chi|\sum_{\mu=1}^{N_n}\sum_{\nu\neq\mu}\sum_{GG'}(\partial_{G'_\nu}A_{G_\mu} - \partial_{G_\mu}A_{G'_\nu})\hat{v}_{G'_\nu}\hat{v}_{G_\mu}|\chi\rangle_{\underline{\underline{\mathbf{R}}}} \\ &= \mathrm{Re}\langle\chi|\sum_{\mu=1}^{N_n}\sum_{\nu\neq\mu}\sum_{GG'}\partial_{G'_\nu}A_{G_\mu}[\hat{v}_{G'_\nu},\hat{v}_{G_\mu}]|\chi\rangle_{\underline{\underline{\mathbf{R}}}}.\end{aligned} \tag{S49}$$

Here we have interchanged the indices $G, G'$ and $\mu,\nu$ in the summation in order to derive the last equality. Now we compute the commutator between $\hat{v}_{G'_\nu}$ and $\hat{v}_{G_\mu}$,

$$\begin{aligned}[\hat{v}_{G'_\nu},\hat{v}_{G_\mu}]\chi &= \frac{1}{4M_\mu M_\nu}\Big((-i\partial_{G'_\nu}+A_{G'_\nu})(-i\partial_{G_\mu}+A_{G_\mu}) - (-i\partial_{G_\mu}+A_{G_\mu})(-i\partial_{G'_\nu}+A_{G'_\nu})\Big)\chi \\ &= \frac{i}{4M_\mu M_\nu}\Big(-\partial_{G'_\nu}(A_{G_\mu}\chi) - A_{G'_\nu}\partial_{G_\mu}\chi + \partial_{G_\mu}(A_{G'_\nu}\chi) + A_{G_\mu}\partial_{G'_\nu}\chi\Big) \\ &= \frac{i}{4M_\mu M_\nu}(\partial_{G_\mu}A_{G'_\nu} - \partial_{G'_\nu}A_{G_\mu})\chi.\end{aligned} \tag{S50}$$

This suggests that

$$\langle\chi|\partial_{G'_\nu}A_{G_\mu}[\hat{v}_{G'_\nu},\hat{v}_{G_\mu}]|\chi\rangle_{\underline{\underline{\mathbf{R}}}} = \frac{i}{4M_\mu M_\nu}\langle\chi|\partial_{G'_\nu}A_{G_\mu}(\partial_{G_\mu}A_{G'_\nu} - \partial_{G'_\nu}A_{G_\mu})|\chi\rangle_{\underline{\underline{\mathbf{R}}}} \tag{S51}$$

is purely imaginary. Therefore, the RHS of Eq. (S49) is zero. Hence, Eq. (S47) is true.

### D. Additional remarks on the SEI for $T_{\mu,marg}$

We have established SEIs for the momentum and angular momentum of individual nuclei. Here we remark that the analogous SEI for individual nuclear kinetic energy is not true, i.e.,

$$\frac{dT_{\mu,marg}}{dt} \neq \mathrm{Re}\langle\chi|\hat{\boldsymbol{\mathcal{F}}}_\mu \cdot \hat{\boldsymbol{v}}_\mu|\chi\rangle_{\underline{\underline{\mathbf{R}}}}. \tag{S52}$$

Here $T_{\mu,marg} = \frac{1}{2M_\mu}\langle\chi|(-i\nabla_{\mathbf{R}_\mu} + \boldsymbol{A}_\mu)^2|\chi\rangle_{\underline{\underline{\mathbf{R}}}}$. Replacing $\hat{\boldsymbol{\mathcal{F}}}_\mu$ by $\boldsymbol{E}_\mu$ or $\hat{\boldsymbol{F}}_\mu$ on the RHS of Eq. (S52) does not lead to the correct formula, instead, a correction operator $\hat{\boldsymbol{\mathcal{G}}}_\mu$ needs to be introduced,

$$\frac{dT_{\mu,marg}}{dt} = \mathrm{Re}\langle\chi|\hat{\boldsymbol{\mathcal{F}}}_\mu \cdot \hat{\boldsymbol{v}}_\mu + \hat{\boldsymbol{\mathcal{G}}}_\mu|\chi\rangle_{\underline{\underline{\mathbf{R}}}}. \tag{S53}$$

By some algebra, one can work out the expression of $\hat{\boldsymbol{\mathcal{G}}}_\mu$ as the following,

$$\hat{\boldsymbol{\mathcal{G}}}_\mu = -\sum_{\nu\neq\mu}\sum_{GG'} \frac{1}{4M_\nu M_\mu}(\partial^2_{G_\mu G'_\nu}\mathcal{C}^{G'G}_{\nu\mu} + 2\partial_{G_\mu}\mathcal{C}^{G'G}_{\nu\mu}\partial_{G'_\nu} + 2\partial_{G'_\nu}\mathcal{C}^{G'G}_{\nu\mu}\partial_{G_\mu}), \tag{S54}$$

where $\mathcal{C}^{G'G}_{\nu\mu} = \partial_{G'_\nu}A_{G_\mu} - \partial_{G_\mu}A_{G'_\nu}$. Derivation is omitted. The operator $\boldsymbol{\mathcal{G}}_\mu$ has no classical analog. Nevertheless, it is worth noticing that $\boldsymbol{\mathcal{G}}_\mu$ has no collective effect, in the sense that

$$\mathrm{Re}\langle\chi|\sum_{\mu=1}^{N_n}\hat{\boldsymbol{\mathcal{G}}}_\mu|\chi\rangle_{\underline{\underline{\mathbf{R}}}} = 0. \tag{S55}$$

## II. CONNECTION BETWEEN OUR FORCE OPERATOR AND THE CLASSICAL FORCE FUNCTION

### A. Summary of classical force functions derived in previous works

In previous works, for the purpose of solving the nuclear Schrödinger equation, trajectory based methods were introduced. By writing $\chi(\underline{\underline{\mathbf{R}}}, t) = \exp\left\{\frac{i}{\hbar}\tilde{S}(\underline{\underline{\mathbf{R}}}, t)\right\}$ and expanding the complex function $\tilde{S}(\underline{\underline{\mathbf{R}}}, t)$ into power series of $\hbar$, i.e., $\tilde{S}(\underline{\underline{\mathbf{R}}}, t) = \sum_\alpha \hbar^\alpha \tilde{S}_\alpha(\underline{\underline{\mathbf{R}}}, t)$, a canonical momenta associated with the lowest order term $\tilde{S}_0$ was defined, [1, 2]

$$\tilde{\boldsymbol{P}}_\mu(\underline{\underline{\mathbf{R}}}, t) = \nabla_{\mathbf{R}_\mu}\tilde{S}_0(\underline{\underline{\mathbf{R}}}, t) + \boldsymbol{A}_\mu(\underline{\underline{\mathbf{R}}}, t). \tag{S56}$$

Moreover, it has been shown that $\tilde{\boldsymbol{P}}_\mu$ along a classical trajectory satisfies the following Newton's equation, [1, 2]

$$\frac{d\tilde{\boldsymbol{P}}_\mu(\underline{\underline{\mathbf{R}}}^I(t), t)}{dt} = \tilde{\boldsymbol{F}}_\mu(\underline{\underline{\mathbf{R}}}^I(t), t). \tag{S57}$$

Here

$$\tilde{\boldsymbol{F}}_\mu = \partial_t \boldsymbol{A}_\mu - \nabla_{\mathbf{R}_\mu}\epsilon - \tilde{\boldsymbol{v}}_\mu \times \boldsymbol{B}_{\mu\mu} + \sum_{\nu\neq\mu} \boldsymbol{F}_{\mu\nu}(\tilde{\boldsymbol{v}}_\nu), \tag{S58}$$

where $\tilde{\boldsymbol{v}}_\mu = \frac{\tilde{\boldsymbol{P}}_\mu}{M_\mu}$, $\boldsymbol{B}_{\mu\nu} = \nabla_{\mathbf{R}_\mu} \times \boldsymbol{A}_\nu$, and

$$\boldsymbol{F}_{\mu\nu}(\tilde{\boldsymbol{v}}_\nu) = -\tilde{\boldsymbol{v}}_\nu \times \boldsymbol{B}_{\mu\nu} + \Big[(\tilde{\boldsymbol{v}}_\nu \cdot \nabla_{\mathbf{R}_\nu})\boldsymbol{A}_\mu - (\tilde{\boldsymbol{v}}_\nu \cdot \nabla_{\mathbf{R}_\mu})\boldsymbol{A}_\nu\Big]. \tag{S59}$$

If one considers all orders of $\hbar$ through the following polar representation of $\chi$, $\chi(\underline{\underline{\mathbf{R}}}, t) = |\chi(\underline{\underline{\mathbf{R}}}, t)|\exp\Big\{\frac{i}{\hbar}S(\underline{\underline{\mathbf{R}}}, t)\Big\}$, the corresponding canonical momenta [3, 4] and force functions associated with $S(\underline{\underline{\mathbf{R}}}, t)$ can be defined as

$$\boldsymbol{P}_\mu(\underline{\underline{\mathbf{R}}}, t) = \nabla_{\mathbf{R}_\mu} S(\underline{\underline{\mathbf{R}}}, t) + \boldsymbol{A}_\mu(\underline{\underline{\mathbf{R}}}, t), \tag{S60}$$

$$\boldsymbol{F}_\mu(\underline{\underline{\mathbf{R}}}, t) = \partial_t \boldsymbol{A}_\mu - \nabla_{\mathbf{R}_\mu}\epsilon - \boldsymbol{v}_\mu \times \boldsymbol{B}_{\mu\mu} + \sum_{\nu\neq\mu} \boldsymbol{F}_{\mu\nu}(\boldsymbol{v}_\nu), \tag{S61}$$

where $\boldsymbol{v}_\mu = \frac{\boldsymbol{P}_\mu}{M_\mu}$. When we consider the total time derivative of $\boldsymbol{P}_\mu$ along a classical trajectory, however, the analogous equality of Eq. (S57) is not true, i.e.,

$$\frac{d\boldsymbol{P}_\mu(\underline{\underline{\mathbf{R}}}^I(t), t)}{dt} \neq \boldsymbol{F}_\mu(\underline{\underline{\mathbf{R}}}^I(t), t). \tag{S62}$$

As is known, a quantum potential correction is needed in order to propagate the equation of motion of $\boldsymbol{P}_\mu$ correctly. [5, 6]

As an additional remark, we note that Eq. (S57) is tied to a classical trajectory. If one tries to extend it from a trajectory to all of $\underline{\underline{\mathbf{R}}}$-space and connect the rate of change of $\tilde{\boldsymbol{P}}_\mu(\underline{\underline{\mathbf{R}}}, t)$ with $\tilde{\boldsymbol{F}}_\mu(\underline{\underline{\mathbf{R}}}, t)$, where $\underline{\underline{\mathbf{R}}}$ and $t$ are independent variables, then an equality cannot be proved, i.e., $\partial_t\tilde{\boldsymbol{P}}_\mu(\underline{\underline{\mathbf{R}}}, t) \neq \tilde{\boldsymbol{F}}_\mu(\underline{\underline{\mathbf{R}}}, t)$. The same argument applies to $\boldsymbol{P}_\mu(\underline{\underline{\mathbf{R}}}, t)$, $\partial_t\boldsymbol{P}_\mu(\underline{\underline{\mathbf{R}}}, t) \neq \boldsymbol{F}_\mu(\underline{\underline{\mathbf{R}}}, t)$. Nevertheless, we have shown in this paper that an equality can be established between $\partial_t\langle\chi|\hat{\boldsymbol{p}}_\mu|\chi\rangle$ and $\mathrm{Re}\langle\chi|\hat{\boldsymbol{\mathcal{F}}}_\mu|\chi\rangle$.

### B. Relation between operators $\hat{\boldsymbol{p}}_\mu$, $\hat{\boldsymbol{\mathcal{F}}}_\mu$ and auxiliary functions $\boldsymbol{P}_\mu$, $\boldsymbol{F}_\mu$

Let us calculate $\frac{\hat{\boldsymbol{p}}_\mu\chi}{\chi}$ using the polar representation of $\chi$.

$$\begin{aligned}\frac{\hat{\boldsymbol{p}}_\mu\chi}{\chi} &= \frac{1}{|\chi|e^{iS}}\Big[(-i\nabla_{\mathbf{R}_\mu} + \boldsymbol{A}_\mu)(|\chi|e^{iS})\Big] = \frac{1}{|\chi|e^{iS}}\Big[-ie^{iS}\nabla_{\mathbf{R}_\mu}|\chi| + |\chi|e^{iS}\nabla_{\mathbf{R}_\mu}S + \boldsymbol{A}_\mu|\chi|e^{iS}\Big] \\ &= \boldsymbol{P}_\mu(\underline{\underline{\mathbf{R}}}, t) - i\frac{\nabla_{\mathbf{R}_\mu}|\chi|}{|\chi|}.\end{aligned} \tag{S63}$$

Therefore, $\boldsymbol{P}_\mu(\underline{\underline{\mathbf{R}}},t) = \mathrm{Re}\frac{\hat{\boldsymbol{p}}_\mu\chi}{\chi}$. This also implies $\boldsymbol{v}_\mu(\underline{\underline{\mathbf{R}}},t) = \mathrm{Re}\frac{\hat{\boldsymbol{v}}_\mu\chi}{\chi}$.

To see the relation between $\hat{\boldsymbol{\mathcal{F}}}_\mu$ and $\boldsymbol{F}_\mu(\underline{\underline{\mathbf{R}}},t)$, we rewrite $\boldsymbol{F}_{\mu\nu}$ in terms of Berry curvatures. Taking the $Z$ component of $\boldsymbol{F}_{\mu\nu}$, we have

$$F^Z_{\mu\nu} = (\boldsymbol{B}_{\mu\nu}\times\boldsymbol{v}_\nu)_Z + \Big[(\boldsymbol{v}_\nu\cdot\nabla_{\mathbf{R}_\nu})A_{Z_\mu} - (\boldsymbol{v}_\nu\cdot\nabla_{\mathbf{R}_\mu})A_{Z_\nu}\Big]. \tag{S64}$$

Here

$$\begin{aligned}(\boldsymbol{B}_{\mu\nu}\times\boldsymbol{v}_\nu)_Z &= B^X_{\mu\nu}v_{Y_\nu} - B^Y_{\mu\nu}v_{X_\nu} = (\nabla_{\mathbf{R}_\mu}\times\boldsymbol{A}_\nu)_X v_{Y_\nu} - (\nabla_{\mathbf{R}_\mu}\times\boldsymbol{A}_\nu)_Y v_{X_\nu}\\ &= (\partial_{Y_\mu}A_{Z_\nu} - \partial_{Z_\mu}A_{Y_\nu})v_{Y_\nu} - (\partial_{Z_\mu}A_{X_\nu} - \partial_{X_\mu}A_{Z_\nu})v_{X_\nu}.\end{aligned} \tag{S65}$$

Substituting Eq. (S65) into Eq. (S64), we have

$$\begin{aligned}F^Z_{\mu\nu} &= (\partial_{Y_\mu}A_{Z_\nu} - \partial_{Z_\mu}A_{Y_\nu})v_{Y_\nu} - (\partial_{Z_\mu}A_{X_\nu} - \partial_{X_\mu}A_{Z_\nu})v_{X_\nu}\\ &\quad + (v_{X_\nu}\partial_{X_\nu} + v_{Y_\nu}\partial_{Y_\nu} + v_{Z_\nu}\partial_{Z_\nu})A_{Z_\mu} - (v_{X_\nu}\partial_{X_\mu} + v_{Y_\nu}\partial_{Y_\mu} + v_{Z_\nu}\partial_{Z_\mu})A_{Z_\nu}\\ &= (\partial_{X_\nu}A_{Z_\mu} - \partial_{Z_\mu}A_{X_\nu})v_{X_\nu} + (\partial_{Y_\nu}A_{Z_\mu} - \partial_{Z_\mu}A_{Y_\nu})v_{Y_\nu} + (\partial_{Z_\nu}A_{Z_\mu} - \partial_{Z_\mu}A_{Z_\nu})v_{Z_\nu}\\ &= \sum_{G'}(\partial_{G'_\nu}A_{Z_\mu} - \partial_{Z_\mu}A_{G'_\nu})v_{G'_\nu}\\ &= \sum_{G'}\mathcal{C}^{G'Z}_{\nu\mu}v_{G'_\nu},\end{aligned} \tag{S66}$$

where in the last line we used the definition of the Berry curvature $\mathcal{C}^{G'G}_{\nu\mu}$ from the main text. Similarly, we can derive the $X$ and $Y$ components of $\boldsymbol{F}_{\mu\nu}$ and summarize as follows,

$$F^G_{\mu\nu} = \sum_{G'}\mathcal{C}^{G'G}_{\nu\mu}v_{G'_\nu}, \qquad (G = X, Y, Z). \tag{S67}$$

Similar expressions have appeared in our operator $\hat{\boldsymbol{\mathcal{D}}}_\mu$, where $v_{G_\nu}$ is replaced by $\hat{v}_{G_\nu}$. This is also reflected in the similarity between the force function $\boldsymbol{F}_\mu(\underline{\underline{\mathbf{R}}},t)$ and our force operator $\hat{\boldsymbol{\mathcal{F}}}_\mu$. By promoting all the velocity functions to velocity operators in the force function, one can recover our force operator. Since $\mathrm{Re}\frac{\hat{\boldsymbol{v}}_\mu\chi}{\chi} = \boldsymbol{v}_\mu(\underline{\underline{\mathbf{R}}},t)$, one can immediately see that $\mathrm{Re}\frac{\hat{\boldsymbol{\mathcal{F}}}_\mu\chi}{\chi} = \boldsymbol{F}_\mu(\underline{\underline{\mathbf{R}}},t)$. As a consequence,

$$\begin{aligned}\mathrm{Re}\langle\chi|\hat{\boldsymbol{\mathcal{F}}}_\mu|\chi\rangle &= \mathrm{Re}\int\chi^*\hat{\boldsymbol{\mathcal{F}}}_\mu\chi d\underline{\underline{\mathbf{R}}} = \mathrm{Re}\int|\chi|^2\frac{\chi^*\hat{\boldsymbol{\mathcal{F}}}_\mu\chi}{\chi^*\chi}d\underline{\underline{\mathbf{R}}}\\ &= \int|\chi|^2\mathrm{Re}\frac{\hat{\boldsymbol{\mathcal{F}}}_\mu\chi}{\chi}d\underline{\underline{\mathbf{R}}} = \int|\chi|^2\boldsymbol{F}_\mu(\underline{\underline{\mathbf{R}}},t)d\underline{\underline{\mathbf{R}}}.\end{aligned} \tag{S68}$$

Thus, the expectation value of $\hat{\boldsymbol{\mathcal{F}}}_\mu$ can be evaluated by replacing the operator by $\boldsymbol{F}_\mu(\underline{\underline{\mathbf{R}}}, t)$. Similarly, we can show that $\langle\chi|\hat{\boldsymbol{p}}_\mu|\chi\rangle = \int |\chi|^2 \boldsymbol{P}_\mu(\underline{\underline{\mathbf{R}}}, t) d\underline{\underline{\mathbf{R}}}$. Therefore, our SEI for the nuclear momentum implies that

$$\partial_t \int |\chi|^2 \boldsymbol{P}_\mu(\underline{\underline{\mathbf{R}}}, t) d\underline{\underline{\mathbf{R}}} = \int |\chi|^2 \boldsymbol{F}_\mu(\underline{\underline{\mathbf{R}}}, t) d\underline{\underline{\mathbf{R}}}. \tag{S69}$$

Finally, we note that our force operator $\hat{\boldsymbol{\mathcal{F}}}_\mu$ only depends on the conditional electronic wave function $\Phi$ and is independent of the nuclear wave function $\chi$, in contrast with the force function $\boldsymbol{F}_\mu(\underline{\underline{\mathbf{R}}}, t)$ that depends on $\chi$ through $\boldsymbol{v}_\mu(\underline{\underline{\mathbf{R}}}, t)$.

## III. SOME DETAILS ABOUT THE EXACTLY SOLVABLE MODEL

### A. Model construction

Let us consider the following time-dependent Schrödinger equation (TDSE) for two nuclei in 1D,

$$i\partial_t \chi = \frac{1}{2M} \sum_{j=1}^{2} \Big( - i\partial_{X_j} + A_j(X_1, X_2, t) \Big)^2 \chi(X_1, X_2, t) + \epsilon(X_1, X_2, t)\chi(X_1, X_2, t). \tag{S70}$$

We aim at finding a nice analytical nuclear wave function $\chi(X_1, X_2, t)$ such that the corresponding scalar and vector potentials are all analytical functions so that our model can be regarded as "exactly solvable". Such choice is certainly not unique. Yet it suffices for our purpose to find one such example to prove our concept.

Noticing the resemblance of Eq. (S70) with a TDSE for 1 nucleus in 2D, let us denote $\mathbf{R} = (X_1, X_2)$ and $\nabla = (\partial_{X_1}, \partial_{X_2})$ so that we can rewrite Eq. (S70) into the following compact form,

$$i\partial_t \chi = \frac{1}{2M} \Big( - i\nabla + \boldsymbol{A}(\mathbf{R}, t) \Big)^2 \chi + \epsilon(\mathbf{R}, t)\chi. \tag{S71}$$

Let $\chi = e^{\beta_1 + i\beta_2}$ with $\beta_1$ and $\beta_2$ being real functions. Dividing $\chi$ on both sides of Eq. (S71), and expanding the $(-i\nabla + \boldsymbol{A})^2$ term leads to

$$i\partial_t \ln \chi = \frac{1}{2M\chi} \Big[ - \nabla^2 \chi - 2i\boldsymbol{A} \cdot \nabla \chi - i(\nabla \cdot \boldsymbol{A})\chi + \boldsymbol{A}^2 \chi \Big] + \epsilon. \tag{S72}$$

Using the identity $\frac{\nabla\chi}{\chi} = \nabla \ln \chi$ and $\frac{\nabla^2\chi}{\chi} = \nabla^2 \ln \chi + (\ln \nabla\chi)^2$, we arrive at

$$i\partial_t \ln \chi = \frac{1}{2M} \Big[ - \nabla^2 \ln \chi - (\nabla \ln \chi)^2 - 2i\boldsymbol{A} \cdot \nabla \ln \chi - i(\nabla \cdot \boldsymbol{A}) + \boldsymbol{A}^2 \Big] + \epsilon. \tag{S73}$$

Now substituting $\ln\chi = \beta_1 + i\beta_2$ into Eq. (S73),we have

$$i\partial_t(\beta_1 + i\beta_2) = \frac{1}{2M}\Big[ - \nabla^2(\beta_1 + i\beta_2) - (\nabla\beta_1 + i\nabla\beta_2)^2 \\ - 2i\boldsymbol{A}\cdot\nabla(\beta_1 + i\beta_2) - i(\nabla\cdot\boldsymbol{A}) + \boldsymbol{A}^2\Big] + \epsilon. \tag{S74}$$

Comparing the real and imaginary part of Eq. (S74), we obtain the following two TD equations of $\beta_1$ and $\beta_2$,

$$-\partial_t\beta_2 = \frac{1}{2M}\Big[ - \nabla^2\beta_1 - (\nabla\beta_1)^2 + (\nabla\beta_2)^2 + 2\boldsymbol{A}\cdot\nabla\beta_2 + \boldsymbol{A}^2\Big] + \epsilon, \tag{S75}$$

$$\partial_t\beta_1 = \frac{1}{2M}\Big( - \nabla^2\beta_2 - 2\nabla\beta_1\cdot\nabla\beta_2 - 2\boldsymbol{A}\cdot\nabla\beta_1 - \nabla\cdot\boldsymbol{A}\Big). \tag{S76}$$

Eq. (S76) multiplied by $2\rho \equiv 2|\chi|^2 = 2e^{2\beta_1}$, after simplification, gives the continuity equation,

$$\partial_t\rho = -\frac{1}{M}\nabla\cdot[\rho(\nabla\beta_2 + \boldsymbol{A})]. \tag{S77}$$

Here $\boldsymbol{J} \equiv \frac{1}{M}\rho(\nabla\beta_2 + \boldsymbol{A})$ is the nuclear current density. On the other hand, solving $\epsilon$ in Eq. (S75) leads to

$$\epsilon = -\partial_t\beta_2 + \frac{1}{2M}\Big[\nabla^2\beta_1 + (\nabla\beta_1)^2 - (\nabla\beta_2 + \boldsymbol{A})^2\Big]. \tag{S78}$$

Let us denote $\boldsymbol{S} = \nabla\beta_2 + \boldsymbol{A}$. Then we can rewrite Eq. (S77) in terms of $X_1$ and $X_2$ as

$$\partial_t\rho = -\frac{1}{M}\Big[\partial_{X_1}(\rho S_1) + \partial_{X_2}(\rho S_2)\Big]. \tag{S79}$$

The internuclear Berry curvature can be written in terms of $\boldsymbol{A}$ or $\boldsymbol{S}$ as

$$B_{12} \equiv \partial_{X_1}A_2 - \partial_{X_2}A_1 = \partial_{X_1}S_2 - \partial_{X_2}S_1. \tag{S80}$$

To construct an analytical solution, let us assume $S_1 = S_1(X_2, t)$ and $S_2 = S_2(X_1, t)$ so that Eq. (S79) reduces to

$$\partial_t\rho = -\frac{1}{M}(S_1\partial_{X_1}\rho + S_2\partial_{X_2}\rho). \tag{S81}$$

Dividing by $\rho$ on both sides of Eq. (S81), we arrive at

$$\partial_t\ln\rho = -\frac{1}{M}(S_1\partial_{X_1}\ln\rho + S_2\partial_{X_2}\ln\rho). \tag{S82}$$

Now let us choose

$$\ln\rho = -\frac{1}{\sigma^2}\Big[\Big(X_1 - g_1(t)\Big)^2 + \Big(X_2 - g_2(t)\Big)^2\Big] + C, \tag{S83}$$

so that the nuclear wave density $\rho$ is a Gaussian of a fixed width $\sigma$ moving along a trajectory $(g_1(t), g_2(t))$. Here $C = -\ln(\sigma^2\pi)$ accounts for the normalization of the wave function. Substituting Eq. (S83) into Eq. (S82) leads to

$$2Mg_1'(t)\Big(X_1 - g_1(t)\Big) + 2Mg_2'(t)\Big(X_2 - g_2(t)\Big) = 2S_1\Big(X_1 - g_1(t)\Big) + 2S_2\Big(X_2 - g_2(t)\Big). \tag{S84}$$

By rearranging the terms, we have

$$-\Big(S_1 - Mg_1'(t)\Big)\Big(X_1 - g_1(t)\Big) = \Big(S_2 - Mg_2'(t)\Big)\Big(X_2 - g_2(t)\Big), \tag{S85}$$

$$\Longrightarrow -\frac{S_1 - Mg_1'(t)}{X_2 - g_2(t)} = \frac{S_2 - Mg_2'(t)}{X_1 - g_1(t)}. \tag{S86}$$

By our assumption, the LHS of Eq. (S86) depends on $X_2$ and $t$ while the RHS of Eq. (S86) depends on $X_1$ and $t$. Therefore, we conclude that

$$-\frac{S_1 - Mg_1'(t)}{X_2 - g_2(t)} = \frac{S_2 - Mg_2'(t)}{X_1 - g_1(t)} = f(t), \tag{S87}$$

where $f(t)$ can be arbitrary function of $t$. Then we arrive at

$$S_1 = f(t)\Big(-X_2 + g_2(t)\Big) + Mg_1'(t), \tag{S88}$$

$$S_2 = f(t)\Big(X_1 - g_1(t)\Big) + Mg_2'(t). \tag{S89}$$

In this work, we choose $g_1(t) = a_0 \frac{t}{\sqrt{M}} \cos\frac{t}{\sqrt{M}} + a_1$, $g_2(t) = a_0 \frac{t}{\sqrt{M}} \sin\frac{t}{\sqrt{M}} - a_1$, where $a_0 = 10^{-3}$Bohr, $a_1 = 1$Bohr, $M = 2000m_e$ is roughly the mass of a hydrogen atom and $t$ is in atomic units. Function $f$ is chosen to be $f(t) = a_2 \sin\frac{t}{\sqrt{M}}$ where $a_2 = 0.1$ Bohr. Here we scale the time by $\frac{1}{\sqrt{M}}$ such that the amplitude of the mean force acting on the nuclear center $\bar{\boldsymbol{F}} = M(\ddot{g_1}, \ddot{g_2})$ is not proportional to $M$. Now the remaining degree of freedom is the phase factor $\beta_2(X_1, X_2, t)$, which determines the gauge. In this work we set it to be zero. As a result, $\boldsymbol{A} = \boldsymbol{S}$ and we summarize the variables in our model in the following,

$$\chi(X_1, X_2, t) = \frac{1}{\sigma\sqrt{\pi}}\exp\Big\{-\frac{1}{2\sigma^2}\sum_{j=1}^{2}\Big(X_j - g_j(t)\Big)^2\Big\}, \tag{S90}$$

$$A_1(X_1, X_2, t) = f(t)\Big(-X_2 + g_2(t)\Big) + Mg_1'(t), \tag{S91}$$

$$A_2(X_1, X_2, t) = f(t)\Big(X_1 - g_1(t)\Big) + Mg_2'(t), \tag{S92}$$

$$\epsilon(X_1, X_2, t) = \frac{1}{2M}\Big(\frac{\nabla^2\chi}{\chi} - A_1^2 - A_2^2\Big). \tag{S93}$$

### B. Supplemental results on $F_\mu$ and $D_\mu$

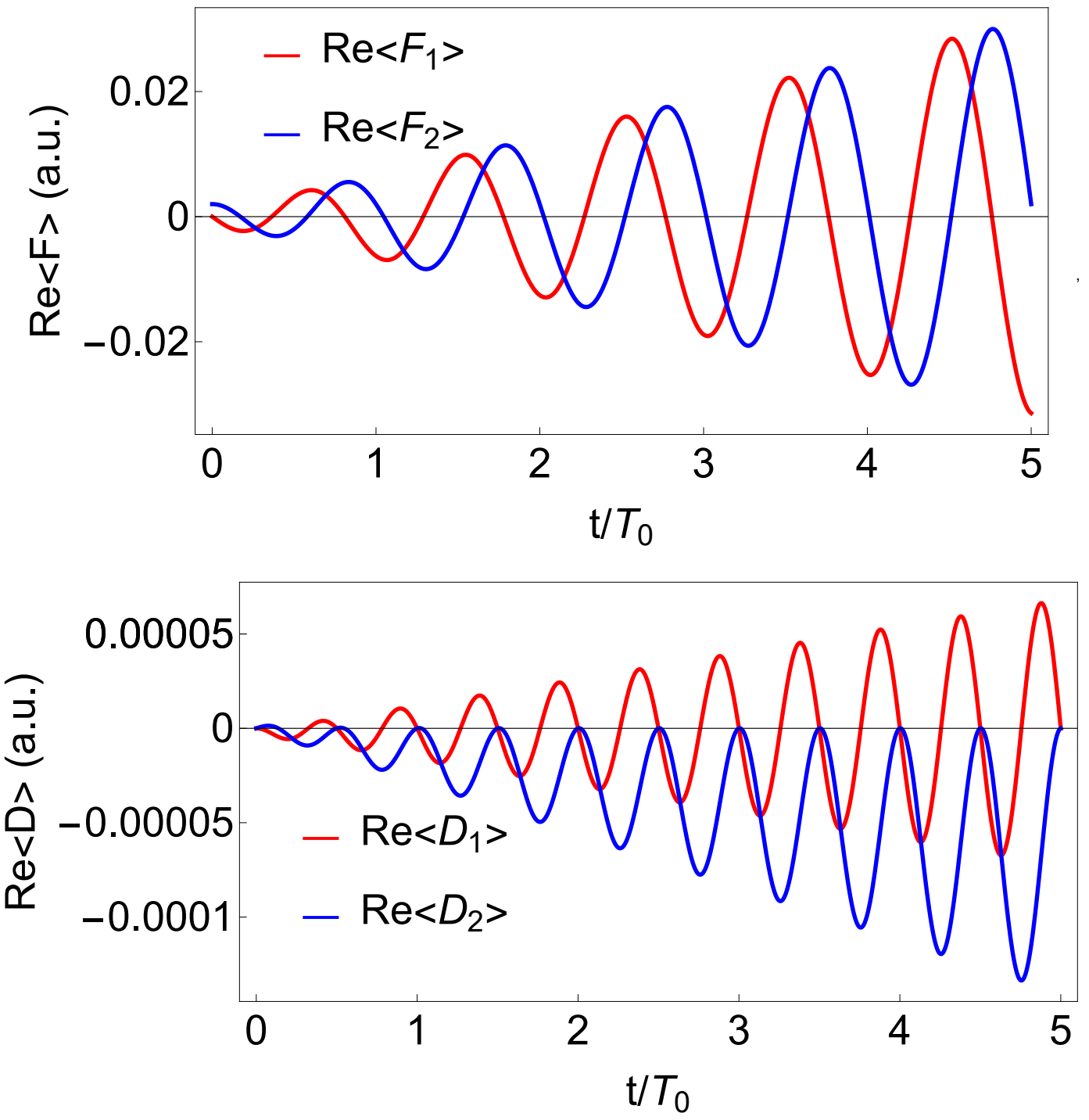

FIG. S1: Comparing $\bar{F}_\mu = \mathrm{Re}\langle\chi|F_\mu|\chi\rangle$ and $\bar{D}_\mu = \mathrm{Re}\langle\chi|D_\mu|\chi\rangle$ as functions of $t$. Here $T_0 = 2\pi\sqrt{M}$ is the time period.

Let us define $\bar{F}_\mu = \mathrm{Re}\langle\chi|\hat{F}_\mu|\chi\rangle$ and $\bar{D}_\mu = \mathrm{Re}\langle\chi|\hat{D}_\mu|\chi\rangle$. By straightforward calculation, for our model we have

$$\bar{F}_1 = Mg_1''(t) + 2f(t)g_2'(t), \quad \bar{D}_1 = -2f(t)g_2'(t), \tag{S94}$$

$$\bar{F}_2 = Mg_2''(t) - 2f(t)g_1'(t), \quad \bar{D}_2 = 2f(t)g_1'(t). \tag{S95}$$

Therefore, the exact mean force acting on atom $\mu$ is given by $\bar{F}_\mu + \bar{D}_\mu = Mg_\mu''(t)$, which generates the correct trajectory $g_\mu(t)$. In Fig. S1, we compare $\bar{F}_\mu$ with $\bar{D}_\mu$ for our spiral model. As one can see, $\bar{D}_\mu$ are 2-3 orders of magnitude smaller than $\bar{F}_\mu$. However, the cumulative effect of $D_\mu$ in time can be sizable. In the main text, we have shown the trajectory led by $\bar{F}_\mu$, which produces a displacement along the $X_2$ direction of about 0.1 Bohr after 10 cycles. From Fig. S1, we see that this effect is exactly counteracted by $\bar{D}_2$ in the negative $X$ direction.

## C. Connection with the atomic water wheel model

The model described here can also be used to simulate one nucleus moving in two dimensions, i.e., we rename $X_1$ and $X_2$ as $X$ and $Y$ coordinate of the same nucleus. This mimicks the current induced atomic water wheel model established in Ref [7], which has important applications in molecular electronic devices. In doing so, the internuclear Berry curvature of the present model becomes the intranuclear Berry curvature, i.e. the induced magnetic field, of the atomic water wheel model, and the internuclear force $\hat{D}_\mu$ in the present model become the induced magnetic forces. We note that magnetic forces have been considered in Ref [7]. However, by fixing the nearby atoms in Ref [7], the motions of these atoms and the effect on the central atom of interest are neglected. If we allow the nearby atoms to move, their motion will produce $\hat{\boldsymbol{D}}_\mu$ that corrects the more classical-like $\hat{\boldsymbol{F}}_\mu$ and changes the nuclear trajectory. This can affect the stability of the atomic water wheel. Our main text shows the trajectory of the center of the nuclear wave packet subject to the correct and the incomplete force. With the incomplete force, the nuclear center gradually moves away from its initial position, suggesting that the motion is not stable. This is, of course, not true because of the missing ingredient in the force.

As an additional remark, the work done by the electrons to the central nucleus in the atomic water wheel is crucially dependent on its mean force, $\bar{\boldsymbol{F}}_\mu = \mathrm{Re}\langle\chi|\hat{\boldsymbol{F}}_\mu + \hat{\boldsymbol{D}}_\mu|\chi\rangle$. In fact, by straightforward calculation, we can calculate the work through the change of marginal nuclear kinetic energy $T_{n,marg}$,

$$T_{n,marg}(t) - T_{n,marg}(0) = \frac{1}{2M}\sigma^2 f^2(t) + \frac{1}{2}M\bar{\boldsymbol{v}}^2. \tag{S96}$$

In the main text, we have argued that the second term (which we call $T_{\mathrm{cm}}$) on the right hand side of Eq. (S96) should dominate. Moreover, we can calculate $T_{\mathrm{cm}}$ through the following line integral of the mean force,

$$T_{\mathrm{cm}} = \int_{\mathcal{C}} \bar{\boldsymbol{F}} \cdot d\boldsymbol{l}. \tag{S97}$$

Here $\mathcal{C}$ is the trajectory of the central nucleus (or the center of mass). If the trajectory is a spiral around its origin with slow increase of radius over time, we can approximate each cycle as a closed path and invoke the Green's theorem to compute the increase of $T_{\mathrm{cm}}$ per

revolution as

$$\Delta T_{\rm cm} = \oint_{\mathcal{C}} \bar{\boldsymbol{F}} \cdot d\boldsymbol{l} = \iint \nabla \times \bar{\boldsymbol{F}} \cdot d\boldsymbol{S}. \tag{S98}$$

If $\nabla \times \bar{\boldsymbol{F}}$ is a slowly varying function in space, $\Delta T_{\rm cm}$ is approximately given by $(\nabla \times \bar{\boldsymbol{F}})\pi r^2$, where $r$ is the radius of the present cycle. This agrees with the analysis in Ref [7] and implies that the mean force is the key quantity that determines the working efficiency of the atomic water wheel. Here, in defining the curl of $\bar{\boldsymbol{F}}$, we shall reconstruct $\bar{\boldsymbol{F}}$ from a function of $t$ to a function of $X$ and $Y$ using the nuclear trajectory $X = X(t)$, $Y = Y(t)$, i.e. we find the function $\bar{\boldsymbol{F}}(X, Y)$ by extending its function value along a trajectory $\bar{\boldsymbol{F}}(X(t), Y(t)) = \bar{\boldsymbol{F}}(t)$ to its enclosed domain. If we further assume that $\bar{\boldsymbol{F}}(X, Y)$ comes from an auxiliary scalar and vector potential, $\bar{\boldsymbol{F}} = \partial_t \tilde{\boldsymbol{A}} - \nabla \tilde{E}$, then the work by the electrons per atomic revolution is

$$\begin{aligned} W = \Delta T_{\rm cm} &= \iint \nabla \times \bar{\boldsymbol{F}} \cdot d\boldsymbol{S} = \iint \nabla \times (\partial_t \tilde{\boldsymbol{A}} - \nabla \tilde{E}) \cdot d\boldsymbol{S} \\ &= \iint \partial_t \nabla \times \tilde{\boldsymbol{A}} \cdot d\boldsymbol{S} = \iint \partial_t \tilde{\boldsymbol{B}} \cdot d\boldsymbol{S}, \end{aligned} \tag{S99}$$

where $\tilde{\boldsymbol{B}} = \nabla \times \tilde{\boldsymbol{A}}$ is the auxiliary magnetic field. Eq. (S99) is a Faraday type of law, although we shall keep in mind the distinction between the auxiliary magnetic field $\tilde{\boldsymbol{B}}$ with the induced magnetic field $\boldsymbol{B} = \nabla \times \boldsymbol{A}$, their corresponding vector potentials and forces.

## IV. SOME DETAILS ABOUT THE VIBRONIC COUPLING MODEL

We design the following 2-level vibronic Hamiltonian to simulate bond length vibrations after a vertical excitation to an excited electronic state of a diatomic molecule. This can be used to study the early stage of a bond-breaking process [8], particularly the reaction pathways of small molecules [9], or the redistribution of energy of DNA and RNA [10, 11]. Let us assume that in the atomic orbital basis the Born-Oppenheimer (BO) Hamiltonian is given by

$$\hat{H}_{\rm BO} = \begin{pmatrix} H_{11} & H_{12} \\ H_{12} & H_{22} \end{pmatrix}. \tag{S100}$$

Moreover, let us assume that the BO surfaces are harmonic potentials,

$$\epsilon_1(R) = \frac{1}{2} m_e \omega_1^2 R^2, \tag{S101}$$

$$\epsilon_2(R) = \frac{1}{2} m_e \omega_2^2 (R - R_0)^2 + \tau, \tag{S102}$$

where $m_e$ is the electronic mass; $\omega_1 = \frac{3}{10}$ a.u., $\omega_2 = \frac{1}{2}$ a.u., $\tau = \frac{1}{30}$ a.u., and $R_0 = 0.5$ Bohr. We design an avoided crossing between the two BO surfaces at around $R_0$ by prescribing the following BO ground and excited state wave functions: $\Phi^1_{\text{BO}} = (\cos\theta, \sin\theta)$ and $\Phi^2_{\text{BO}} = (-\sin\theta, \cos\theta)$, where

$$\cos\theta = \frac{1}{1 + e^{\alpha(R - R_0)}}, \tag{S103}$$

$$\sin\theta = \sqrt{1 - \cos^2\theta}. \tag{S104}$$

In Figure 2(a) of the main text, we have shown the two BO surfaces. Here in Fig. S2, we show the ground state BO electronic wave function by plotting $\cos^2\theta$ as a function of $R$ for different choices of $\alpha$. As one can see, $\alpha$ reflects the sharpness of the electronic transition at the avoided crossing and thus can be used to tune the nonadiabaticity of the model. In Figure 2 of the main text, we have chosen $\alpha = 10$, corresponding to an intermediate nonadiabatic coupling regime.

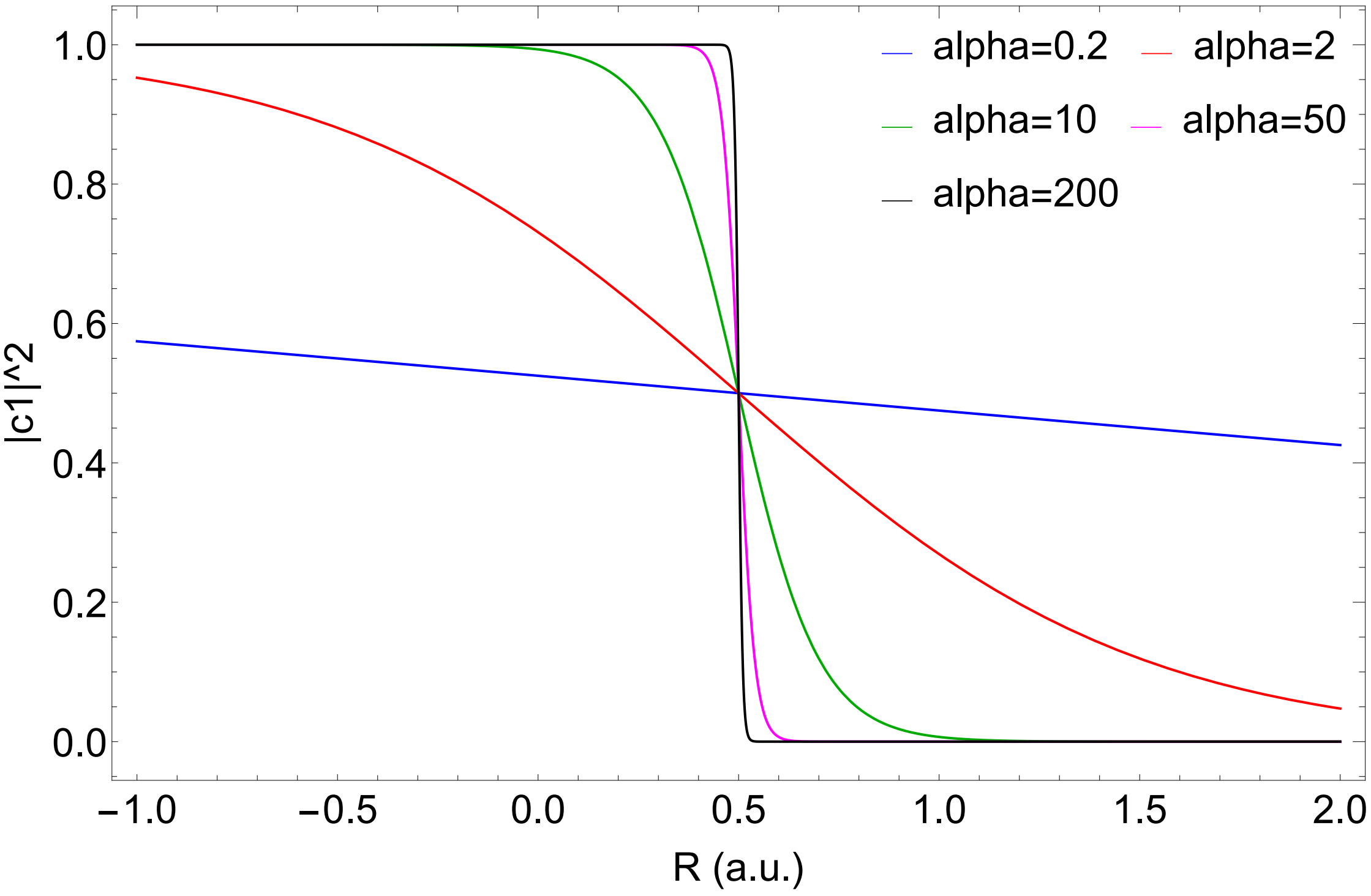

FIG. S2: Comparing $|c_1|^2 = \cos^2\theta$ under BO approximation with different choices of $\alpha$ (in unit of $\text{Bohr}^{-1}$).

Now let us prepare the initial state at the equilibrium bond length of the ground BO

surface, vertically excite the system to its electronic excited state, and then let it propagate. That is, we solve the following time dependent problem:

$$i\partial_t \Psi = \hat{H}\Psi = (-\frac{1}{2M}\frac{d^2}{dR^2} + \hat{H}_{\mathrm{BO}})\Psi, \tag{S105}$$

$$\Psi(t=0) = \Psi_0 = \chi^1_{\mathrm{BO}}(R)\Phi^2_{\mathrm{BO}}(R), \tag{S106}$$

where $\chi^1_{\mathrm{BO}}(R)$ is the lowest eigenstate of $-\frac{1}{2M}\frac{d^2}{dR^2} + \epsilon_1(R)$ and $M = 2000m_e$. We solve the equation by projecting $\hat{H}$ and $\Psi$ onto electron-nuclear basis functions, where we use the BO ground surface harmonic eigenmodes as the nuclear basis functions, i.e.,

$$\chi^1_n(R) = \frac{1}{\sqrt{2^n n!}}\left(\frac{M\omega}{\pi}\right)^{1/4} e^{-\frac{M\omega R^2}{2}} H_n\left(\sqrt{M\omega}R\right). \tag{S107}$$

Here $\omega = \sqrt{\frac{m_e}{M}}\omega_1$ and $H_n$ is the $n$th order Hermite polynomial. Our initial nuclear wave function is $\chi^1_{\mathrm{BO}}(R) = \chi^1_0(R) = \left(\frac{M\omega}{\pi}\right)^{1/4} e^{-\frac{M\omega R^2}{2}}$. In our calculations, we use 32 nuclear basis functions.

Because the total electron-nuclear Hamiltonian is time independent, the total energy of the system is conserved. However, energy can transfer between the electronic and the nuclear subsystems. To see this, one can monitor the nuclear kinetic energy as a function of time.

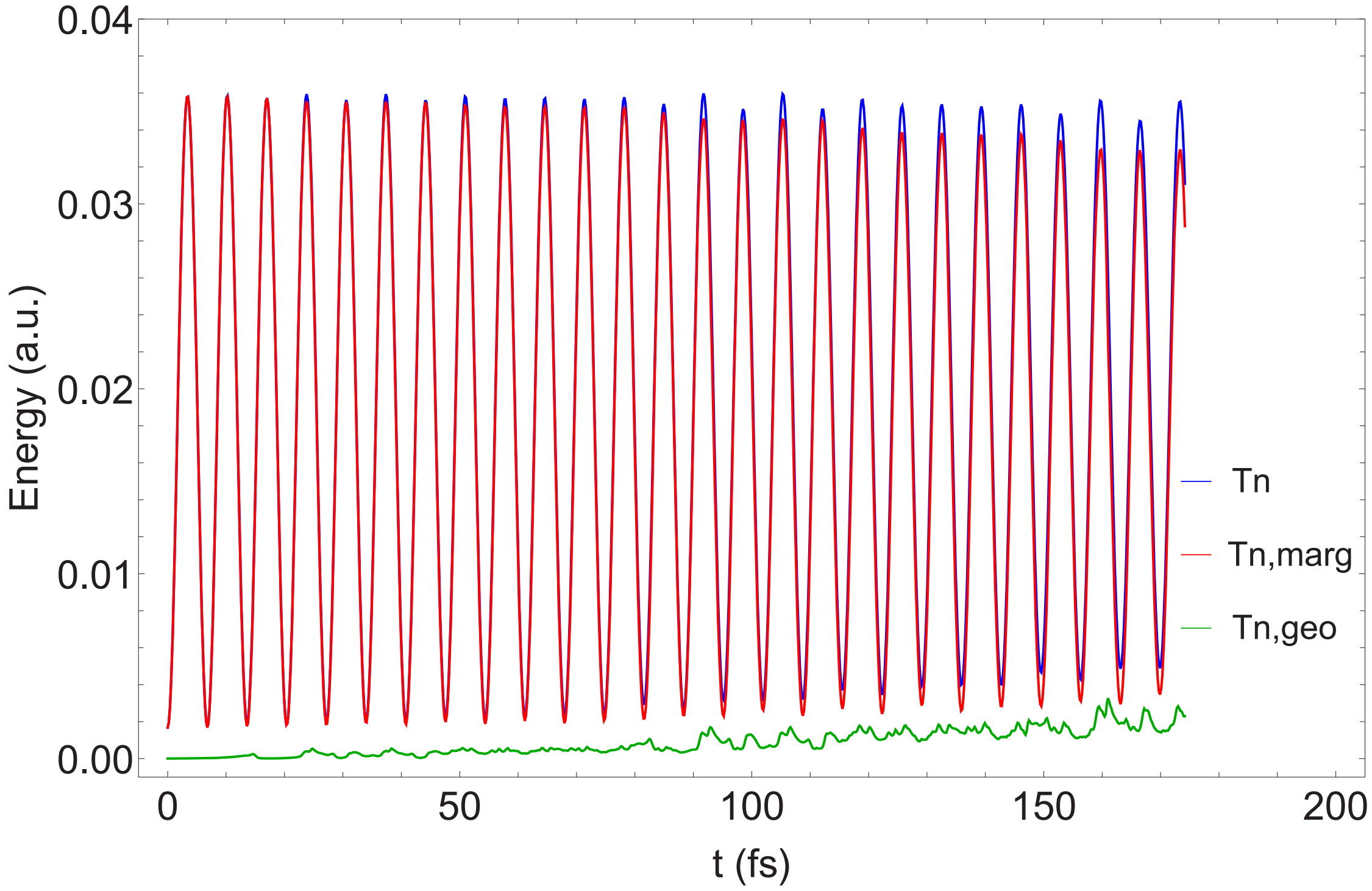

FIG. S3: Comparing $T_n$, $T_{n,marg}$ and $T_{n,geo}$ during time evolution. Here we present the result for $\alpha = 0.4$, which is in the adiabatic regime.

In Fig S3, we show the nuclear kinetic energy $T_n$ and its components ($T_{n,marg}$ and $T_{n,geo}$) for $\alpha = 0.4$ Bohr$^{-1}$, which is in the adiabatic regime. From the figure, we can identify two major features. First, $T_{n,geo}$ is small and $T_{n,marg} \approx T_n$. This allows us to use our Ehrenfest identity to approximately quantify the rate of change of $T_n$. However, it is worth remarking that in the intermediate or nonadiabatic regimes (for example when $\alpha = 10$ Bohr$^{-1}$ or larger), $T_{n,geo}$ is not negligible (figure not shown). Second, as one can see, during time evolution $T_n$ essentially goes through periodic oscillations. In fact, if we turn off the adiabatic coupling, i.e. setting $\alpha = 0$ Bohr$^{-1}$, the nuclear packet will oscillate on the excited BO surface, leading to a perfectly periodic oscillation of $T_n$, with a time period $T_0 = \frac{2\pi}{\omega_2}\sqrt{\frac{m_e}{M}} = 13.6$ fs (note that in each period $T_n$ goes through two maxima, one in right-moving and the other in left-moving). In this trivial case, the nuclear wave packet behaves like a classical pendulum, whose kinetic energy and potential energy continuously transform into each other.

To better understand the energy transfer between electrons and nuclei for finite nonadiabatic coupling, let us define the following running average of $T_n$ over one period:

$$\langle T_n(t) \rangle = \frac{1}{T_0} \int_t^{t+T_0} T_n(t')dt'. \tag{S108}$$

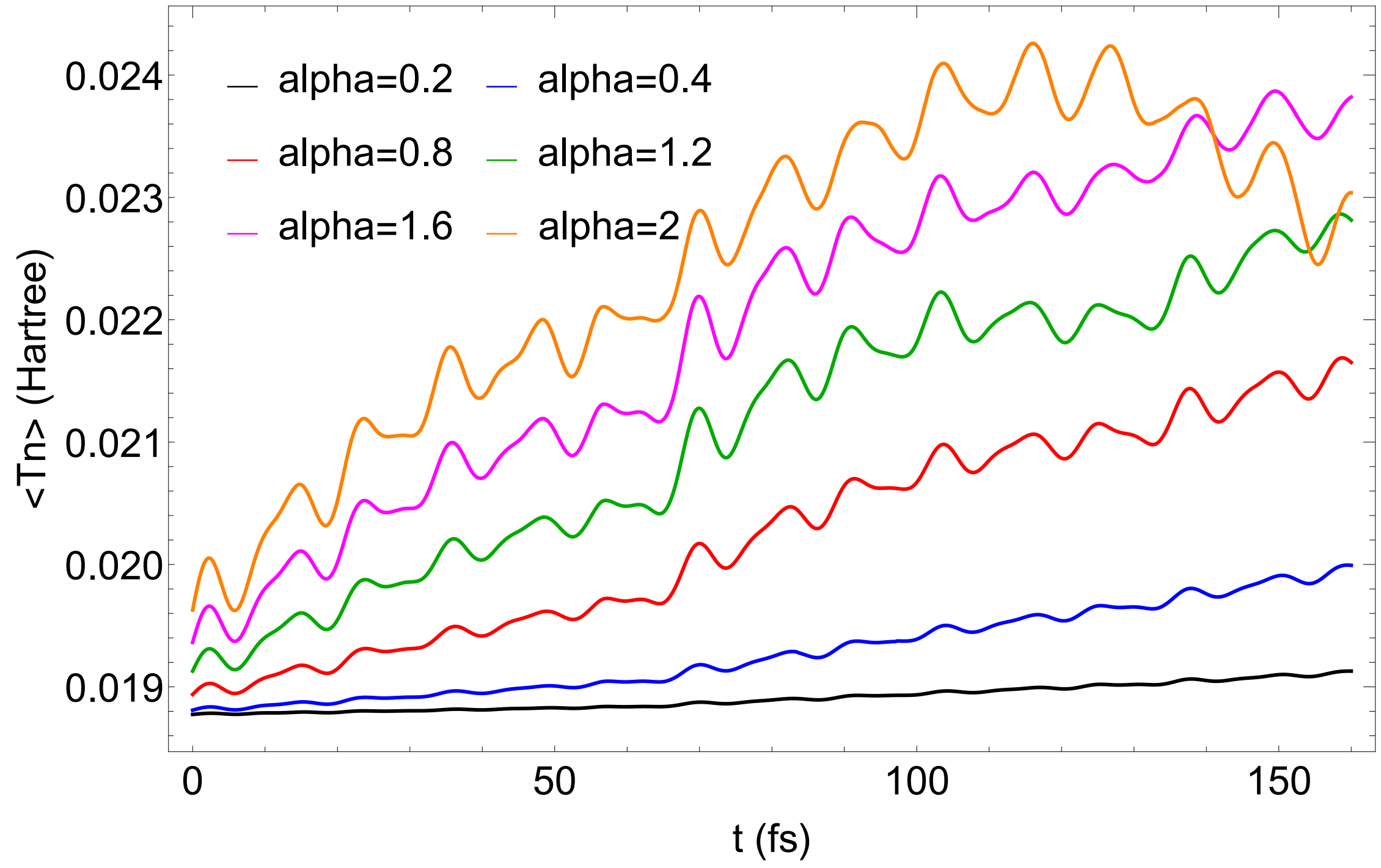

FIG. S4: Running average of $T_n$ in the adiabatic regime.

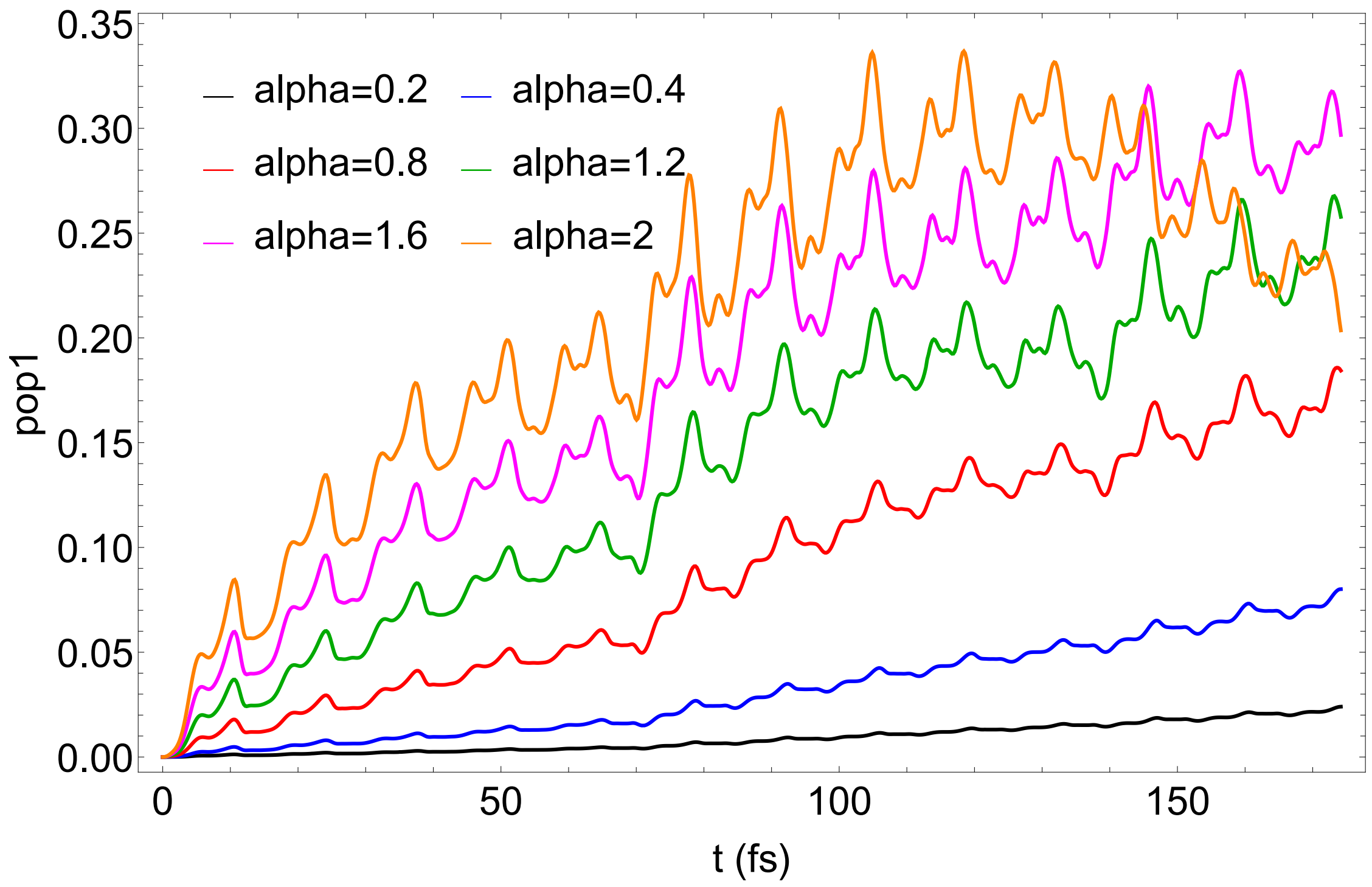

FIG. S5: Electronic population on the ground surface in the adiabatic regime.

We plot $\langle T_n(t)\rangle$ for different choices of $\alpha$ in Fig S4. As we can see, in the adiabatic regime (small $\alpha$) the overall trend of $\langle T_n(t)\rangle$ is to increase in time, suggesting that there is a continuous transfer from the electronic energy to the nuclear kinetic energy. Moreover, this trend is consistent with the steady return of the electronic population to the ground BO surface, as defined by

$$p_1(t) = \int_{-\infty}^{\infty} |\chi(R,t)|^2 |\langle \Phi_{\mathrm{BO}}^1(R)|\Phi(R,t)\rangle|^2 dR. \tag{S109}$$

The plot of $p_1(t)$ is shown in Fig S5. When electrons transfer from the upper surface to the lower surface through the avoided crossing, they lower their electronic energy. This energy is transferred to $T_n$. However, this picture is only valid in the adiabatic regime. As $\alpha$ increases and reaches intermediate or nonadiabatic regimes, the reverse transfer of electronic population from the lower to the upper surface will also become appreciable, leading to the backward transfer of $T_n$ to the electronic energy. Results of these regimes are not shown.

---